\documentclass[useAMS,usenatbib,usegraphicx]{mn2e}

\def\text{\rm}

\newcommand{\cesam}{\textsc{Cesam} }

\newcommand{\PA}[2]{\frac{\partial#1}{\partial#2}}

\newcommand{\FLor}{\textit{\textbf{F}}_{\!\mathbf{{\mathcal{L}}}}}
\newcommand{\bnab}{\mathbf{\nabla}}

\newcommand{\ephi}{\mathbf{\hat{e}_{\varphi}}}
\newcommand{\mbf}[1]{\textbf{\textit{#1}}}
\newcommand{\vphi}{\varphi}

\title{Effect of a fossil magnetic field on the structure of a young Sun }
\author[V. Duez et al.]{%
V. Duez\thanks{vincent.duez@cea.fr},
S. Mathis\thanks{stephane.mathis@cea.fr},
S. Turck-Chi\`eze\thanks{sylvaine.turck-chieze@cea.fr}
\vspace{0.5cm}
\\
DSM/IRFU/SAp, CEA Saclay, 91191 Gif-sur-Yvette Cedex, France;\\ \noindent AIM, UMR 7158, CEA - CNRS - Universit\'e Paris 7, France}

\begin{document}
\maketitle
\begin{abstract}
We study the impact of a fossil magnetic field  on the physical quantities which describe the  structure of a young Sun of 500 Myr. We consider for the first time a non force-free field composed of a mixture of poloidal and toroidal  magnetic fields and we propose a specific configuration to illustrate our purpose. 
In the present paper, we estimate the relative role of the different terms which appear in the modified stellar structure equations. We note  that the Lorentz tension plays a non negligible role in addition to the magnetic pressure. This is interesting because most of the previous stellar evolution codes ignored that term and the geometry of the field. The solar structure perturbations are, as already known, small and consequently we have been able to estimate each term semi-analytically. We develop a general treatment to calculate the global modification of the structure and of  the energetic balance. We estimate also  the gravitational multipolar moments associated with the presence of a fossil large-scale magnetic field in radiative zone. The values given for the young Sun help the future implementation in stellar evolution codes. This work can be repeated for any other field configuration and prepares the achievement of a solar MHD model where we will follow the transport of such field on secular timescales and the associated transport of momentum and chemicals. The described method  will be applied at the present Sun  and the results will be compared with the coming balloon or space measurements.
\end{abstract}
%
%
%
\section{\label{intro} INTRODUCTION}

Stellar magnetic fields, though they are observed more and more extensively at stellar surfaces through spectropolarimetry, remain subtle physical actors in stellar evolution. Intense researches are today devoted to understand their role in convective layers, in particular for the Sun (Cattaneo 1999; Brun et al. 2004; Brun et al. 2005; V\"ogler \& Sch\"ussler 2007; Featherstone et al. 2007) as well as in radiative zones (Charbonneau \& MacGregor 1993; R\"udiger \& Kitchatinov 1997; Spruit 1999-2002; Garaud 2002; Braithwaite \& Spruit 2004; Braithwaite 2006a; Braithwaite \& Nordlund 2006; Brun \& Zahn 2006; Zahn et al. 2007; Garaud \& Garaud 2008), but few work has been devoted to the effects of a fossil field in radiative zone on the stellar structure and its evolution \citep{MM04, eggenberger05}. 

In the previous works, the magnetic field is generally treated rather simply and the geometry totally ignored \citep{lyd95, cou03, ras06}. This fact is due to  the lack of constraints on its magnitude and configuration. Nevertheless, today more activity is focused on the radiative zone \citep{bur04, ras07}, so it becomes useful to improve our approach.

In this context,  2-D models are in construction (see Rieutord 2006, Li et al. 2006) and begin to give some preliminary results for the rotation (Espinosa Lara \& Rieutord 2007)  and for the magnetic field (Li et al. 2008). An alternative  approach resides in the implementation in a unidimensionnal code  of the perturbations of the structure due to the dynamical phenomena. This approach is justified for solar like stars where they are known to be small. In this case, one  considers the projection on the first low-orders spherical harmonics of the equations of the angular momentum transport,  the heat transport and of the induction for the magnetic field  \citep{mat05}. The corresponding dynamical  model contains all the refinements of the microscopic physics present in 1-D stellar evolution codes (see Talon et al. 1997; Meynet \& Maeder 2000; Decressin et al. 2009 and Turck-Chi\`eze et al. 2009 for the case of rotation). 

This paper prepares the implementation in a stellar evolution  code of the different terms induced by the presence of a magnetic field. With this objective, we choose a specific configuration which may result from a fossil field, evaluate the order of magnitude of the different terms induced by the field and estimate the impact of the magnetic field on the structure in developing some semi analytic  calculations well adapted to the small effects that we obtain for  a solar fossil field.\\

Section 2 shows how the stellar structure equations are modified by the presence of a large-scale magnetic field. Then we propose a  relaxed non force-free fossil magnetic field configuration located in the solar radiative zone. It  will be used in a model at its arrival on the main sequence. 
In \S3, the modification of the mechanical balance is studied. Radial perturbations of the gravitational potential, density, pressure, and radius are  computed up to the surface.  The method used is described in {appendix}. Then, in \S4, the perturbation of the energetic balance is examined. We establish the change in temperature owing to the perturbation in density and pressure.  We investigate the energetic perturbations generated by the Ohmic heating, the Poynting's flux, and by the change in nuclear reaction rates induced by the modification of the mechanical balance. In \S5,  we estimate the surface perturbations and the gravitational multipolar moments induced by the presence of the magnetic field. Finally, \S6 summarizes the results and shows the perspectives.

It is important to remark that although our study is  focused on some specific solar fossil field, the formalism that we derive  is general and can be applied to any non-axisymmetric or/and time-dependent magnetic fields.

\section{\label{stellarevo} 
THE DEEP FOSSIL MAGNETIC FIELD}
\subsection{\label{structeq}The Modified Stellar Structure Equations}
We recall first how the stellar structure equations are modified by the presence of a magnetic field $\mbf{B}\left(r,\theta,\varphi,t\right)$:
\begin{equation}
\frac{\partial \left<P_{\rm gas}\right>_{\theta,\varphi}}{\partial M_r} = -\frac {GM_r}{4\pi r^4} +  \frac{1}{4 \pi r^2}\left<\frac{{F}_{\mathcal L; r}}{\rho}\right>_{\theta,\varphi};
\label{eqstructP}
\end{equation}
\begin{equation}
\frac{\partial M_r}{\partial r} = 4\pi r^2\left<\rho\right>_{\theta,\varphi};
\label{eqstructR}\\
\end{equation}
\begin{equation} 
\frac{\partial L}{\partial M_r} = \left< \varepsilon - \frac{\partial U}{\partial t} +
\frac{P_{\rm gas}}{\rho^2}\frac{\partial \rho}{\partial t} + \frac{1}{\rho}Q_{\rm Ohm} +  \frac{1}{\rho} F_{\rm Poynt} \right>_{\theta,\varphi};
\label{eqstructL}\\ 
\end{equation}
\begin{equation}
\frac{\partial \left<T\right>_{\theta,\varphi}}{\partial M_r} =  \frac{\partial P_{\rm gas}}{\partial M_r} \frac{\left<T\right>_{\theta,\varphi}}{P_{\rm gas}} \nabla;
\label{eqstructT}
\end{equation}
$\rho$ being the density, $M_r$ the mass contained in a sphere with a radius $r$, $G$ the gravitational constant, $P_{\rm{gas}}$ the gas pressure, $L$ the luminosity, $\varepsilon$ the energy production rate  {per unit of mass}, $U$ the internal energy {per unit of mass}, $T$ the temperature, $\nabla$ is $\nabla_{rad}$ in radiative zones and $\nabla_{ad}$ in convective zones. The new terms are $P_{\mathrm{mag}} =\mbf{B} ^2 / 2 \mu_0 $, the Lorentz force $ {\mbf{F}}_{\mathcal{L}}  = \mbf{j}  \times\mbf{B} $, the Ohmic heating $Q_{\rm Ohm}= (1/ \mu_0 )\left[||\eta||\otimes\left(\bnab \times\mbf{B} \right)\right] \cdot \left(\bnab \times\mbf{B} \right)$, and the Poynting's flux $F_{\rm Poynt}= (1/ \mu_0) \bnab \cdot \left( \mbf{E}  \times \mbf{B}  \right)$, $\mbf{j} $ being the current density, $||\eta||$ the magnetic diffusivity tensor and $\mu_0$ the vacuum magnetic permeability. $\left(r,\theta,\varphi\right)$ are the usual spherical coordinates.

In a classical stellar evolution code, the equations of the mechanical (eq. \ref{eqstructP}) and the energetic (eq. \ref{eqstructL}) balances are solved only radially. However, when the topology of the magnetic field is introduced, the star becomes three-dimensional. So the multi-dimensional quantities are averaged over the colatitudes ($\theta$) and the azimuthal angle ($\varphi$) according to $\left<Z\right>_{\theta,\varphi}=1/4\pi\int_{0}^{2\pi}\int_{0}^{\pi}Z(r,\theta,\varphi)\sin\theta {\rm d}\theta{\rm d}\varphi$.

\subsection{\label{bgeo} The Magnetic Field Topology}
In the present work, we consider a large-scale magnetic field geometry likely to exist in stellar interiors, especially in radiative zones of solar-like stars on the main sequence. 
\cite{tay73} has shown that purely toroidal fields are unstable (see also Braithwaite 2006b). Moreover, \cite{mar73, mar74} and in parallel,  \cite{wri73} deduced  that purely poloidal fields are also unstable (see also Braithwaite 2007). So, we consider a mixed poloidal-toroidal configuration which can survive over evolution timescales. 

When the Sun arrives on the main sequence, the magnetic field is probably only a perturbation compared with the gravitational potential and the gaseous pressure gradient (high-$\beta$ regime) and the fossil magnetic field are constrained to relax in a non force-free equilibrium.
Hence, we focus on such relaxed fossil configurations 
to predict its influence upon the solar structure. In this first study, we do not consider the  rotation history (see \citep{Turck09}) to isolate the effect of the magnetic field.  The magnetic field is chosen axisymmetric, in a MHS equilibrium as an initial condition and is expressed as a function of a poloidal flux $\Psi(r,\theta)$ and a toroidal potential $F(r,\theta)$ so it remains divergence-free by construction :
\begin{eqnarray}
\mbf{B} =
 \frac {1}{r \sin \theta} \bnab \Psi \mathbf{\times} \ephi + \frac{1}{r \sin \theta}\: F \;\ephi,
\label{bofpsiandf}
\end{eqnarray}
where in spherical coordinates the poloidal direction is in the meridional plane ($\widehat{\mbf e}_r,\widehat{\mbf e}_{\theta}$) and the toroidal direction is along the azimuthal one (along $\widehat{\mbf e}_{\varphi}$), $\left\{\widehat{\mbf e}_k\right\}_{k=r,\theta,\varphi}$ being the spherical coordinates unit vectors basis. 
Without any known constraint on its topology and its strength, we have adopted the following magnetic field's poloidal flux \citep{pre56, woltjer59, woltjer60, wentzel61} :
\begin{eqnarray}
\lefteqn{\Psi\left(r,\theta \right)=}\nonumber\\
&&
\sin^2\theta\,\times \Bigg\{ \sum_{l=0}^{\infty}K_1^l\frac{\lambda_{1}^{l,i}}{R_{\rm sup}}\,r\, j_{l+1} \left(\lambda_1^{l,i}\frac{\: r}{R_{\rm sup}}\right) C_{l}^{3/2}\left(\cos \theta\right)
\nonumber\\
&-&\!
\mu_0\beta_0\frac{\lambda_{1}^{0,i}}{R_{\rm sup}}r j_{1}\!\!\left(\lambda_{1}^{0,i}\frac{r}{R_{\rm sup}}\right)\!\int_{r}^{R_{\rm sup}}\!\left[\,y_{1}\!\!\left(\lambda_{1}^{0,i}\,\frac{\xi}{R_{\rm sup}}\right)\overline\rho\,\xi^3\!\right]\!{\rm d}\xi
\nonumber\\
&-&\!
\mu_0\beta_0\frac{\lambda_{1}^{0,i}}{R_{\rm sup}}r y_{1}\!\!\left(\lambda_{1}^{0,i}\,\frac{r}{R_{\rm sup}}\right)\!\int_{R_{\rm inf}}^{r}\!\left[
\,j_{1}\!\!\left(\lambda_{1}^{0,i}\,\frac{\xi}{R_{\rm sup}}\right)\!\overline\rho\,\xi^3\!\right]\!{\rm d}\xi\!
\Bigg\}\nonumber\\
\end{eqnarray}
where $\overline\rho$ is the non-magnetic equilibrium density, $j_l$ and $y_l$ being respectively the spherical Bessel and Neumann functions of latitudinal order $l$.  The $C_{l}^{3/2}$ are the classical Gegenbauer functions. The eigenvalues $\lambda_1^{l,i}$ ($i \geq 1$) and the constants $K_1^l$ ($l \geq 0$) are given by the boundary conditions at the bottom of the considered radiation zone (at $r=R_{\rm inf}$) and at its upper limit (at $r=R_{\rm sup}$) and the parameter $\beta_0$ is constrained by the magnetic field strength. The  chosen boundary conditions are $\left(\mbf B\cdot\widehat{\mbf e}_r\right)=0$ $\left(\Psi =0\right)$ at the radiation-convection zones interfaces (i.e. at $r=R_{\rm inf/sup}$) and at the center for the solar case. We focus here on the dipolar mode l=1.

Then we choose to derive the toroidal potential in the following manner
\begin{eqnarray}
F \left(\Psi \right) = \lambda_1^0\: \Psi.
\end{eqnarray}
\begin{figure}
\begin{center}
\includegraphics[width=6cm]{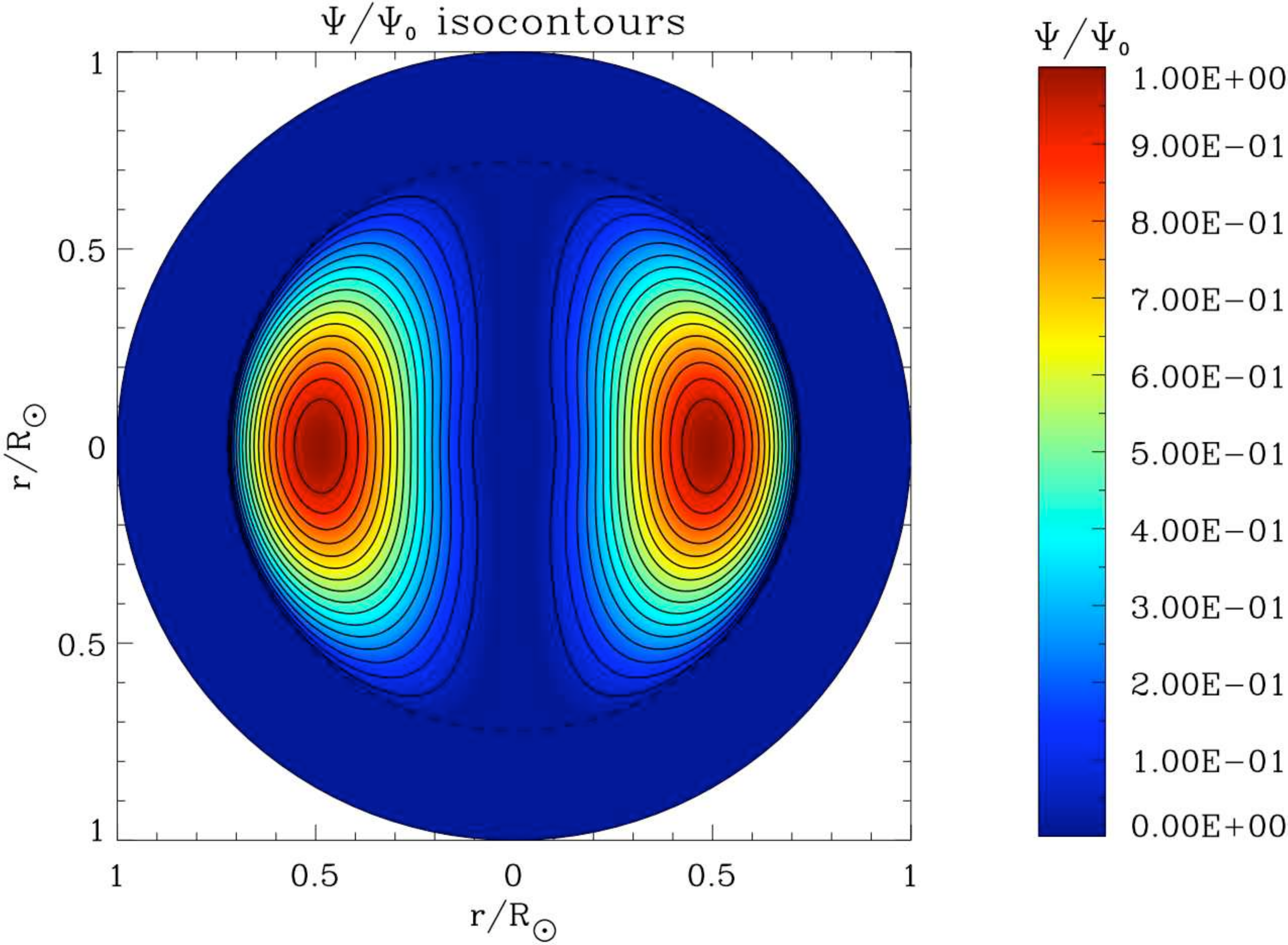}
\includegraphics[width=6cm]{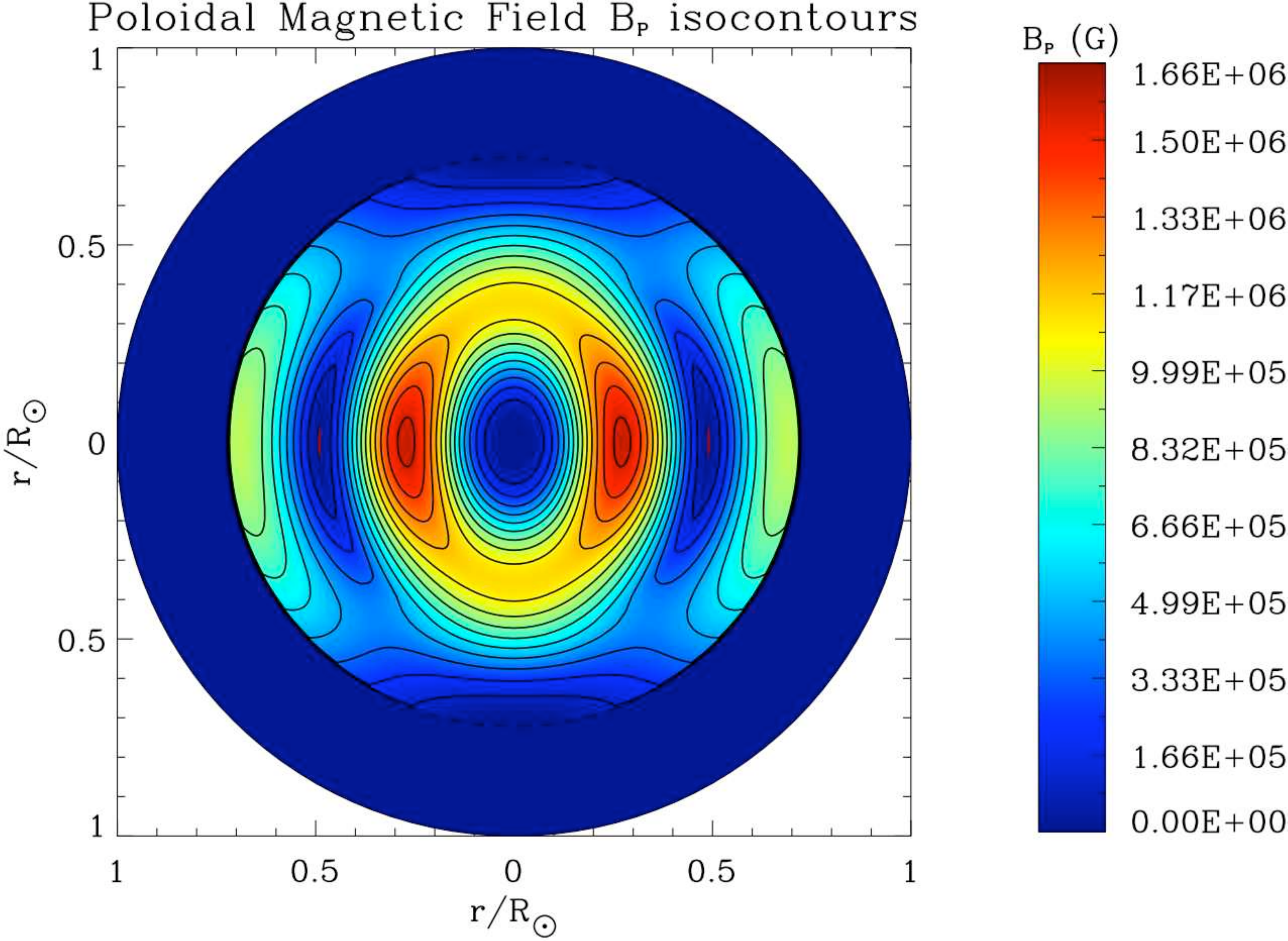}
\includegraphics[width=60mm]{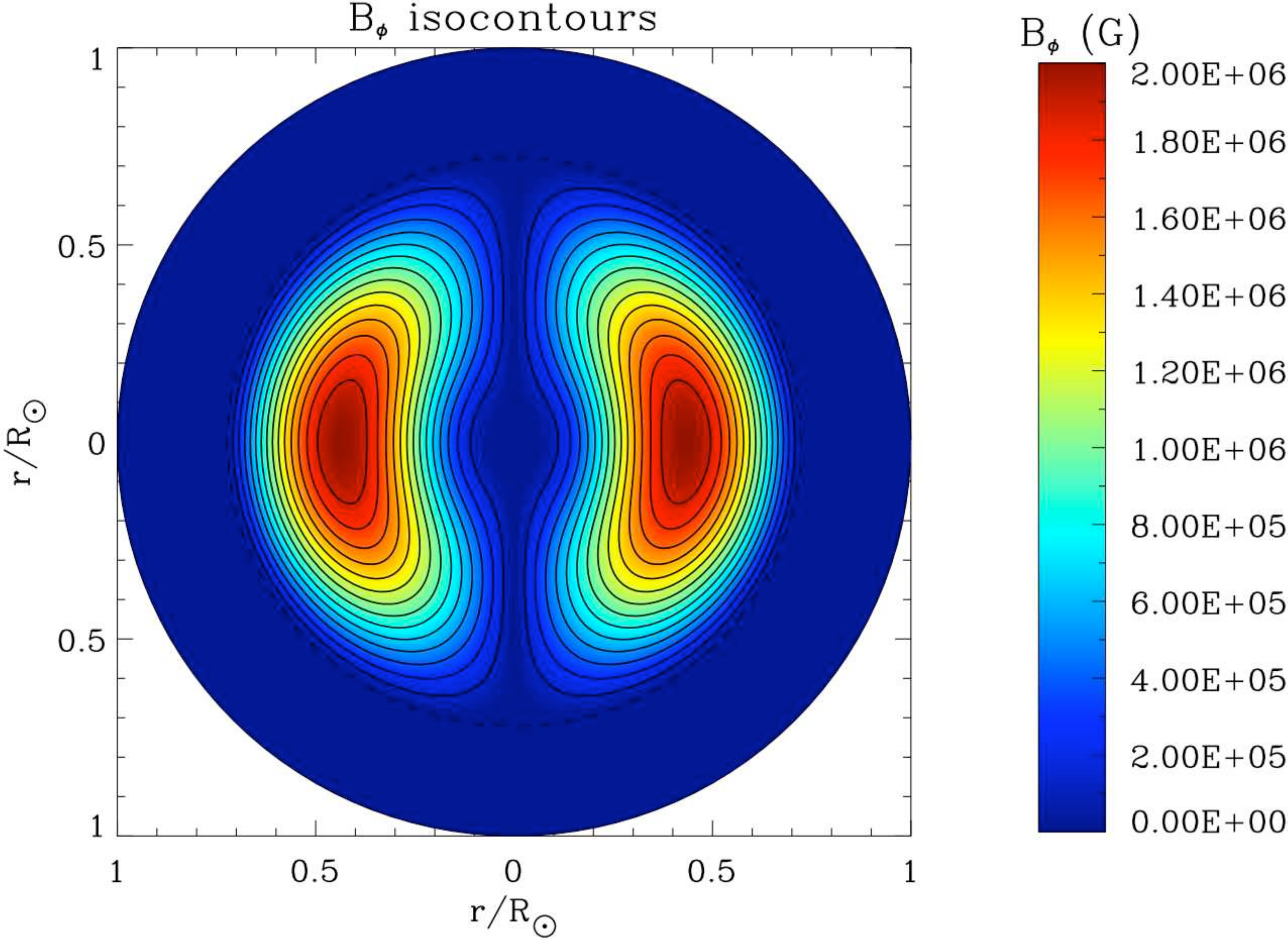}
\caption{a) Isocontours of the flux function $\Psi(r, \theta)$ normalized to its maximum in meridional cut. 
The poloidal magnetic field is tangent to the iso-contours. b) Isocontours of the poloidal magnetic field $B_\varphi(r, \theta)$ (in G) in meridional cut.
c) Isocontours of the azimuthal magnetic field $B_\varphi(r, \theta)$ (in G) in meridional cut.
\label{psi}}
\end{center}
\end{figure}
\begin{figure}
\begin{center}
\includegraphics[width=0.4\textwidth]{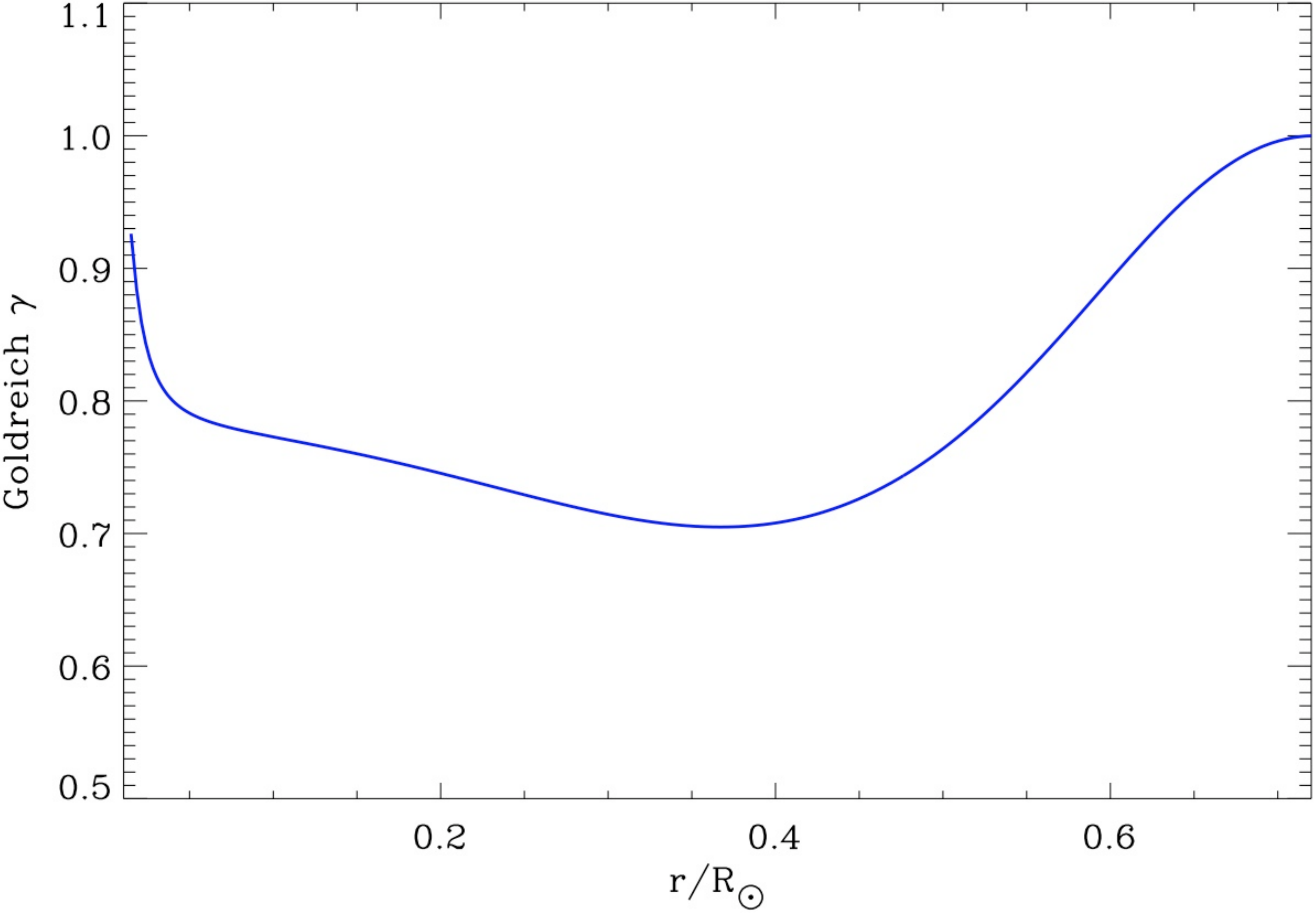}
\caption{Golreich's parameter, 
In this application one supposes a mixed field of 2 MG located around 0.3 R$\odot$.
\label{Goldreich-Beta}}
\end{center}
\end{figure}
The field is buried in the radiative zone. Of course, this field will tend to diffuse outwards due to Ohmic dissipation, as observed in the numerical experiment performed by Braithwaite \& Nordlund (2006). It will also be modified by the meridional circulation and the differential rotation. {This is why we choose here to focus on its impact on a young Sun.}

\subsection{The solar model and the related magnetic field strength}
We first calculate the solar structure using the \cesam code \citep{morel97}.
We consider a non-rotating model at the arrival on the main sequence with initial conditions  determined for the solar seismic model following \citet{cou03} and \citet{tur04}. It was computed using  \cite{gn93} abundances, taking into account the microscopic diffusion according to the \cite{mp93} formalism, and using \textsc{Opal} equation of state  and opacity tables (Rogers, Swenson \& Iglesias 1996; Iglesias \& Rogers 1996). The initial hydrogen fraction is $X_0 = 0.7001$, the initial helium one is $Y_0 = 0.272$. We let this model evolved up to 500 Myr. At this age, the radius (here called $R_{\odot}$ in the figures) is 0.886 of the present solar radius and the luminosity is 76.2 \% of the present solar luminosity.

In this study, we consider  a magnetic field strength  of $B_0 = 2 \: {\rm MG}$  at its maximum \citep{fri04}, located around $0. 35 {\rm R}_{\odot}$. 
This is a pure example and other cases can be studied  if we discover good arguments to prefer some other topologies and strengths. These initial values will evolve with time but we need to determine them  to estimate the momentum transport by rotation and magnetic field. The chosen  isocontours of the function $\Psi$  and the resulting poloidal $B_{\rm P}$ and azimuthal magnetic fields $B_{\vphi}$ are plotted in Fig. \ref{psi}. 
As shown by Goldreich (1991), an anisotropy factor $\gamma_{\rm G}$ can be defined as 
\begin{eqnarray}
\gamma_{\rm G} = \frac{<B_{\rm h}^2>-<B_{\rm r}^2>}{<B_{\rm h}^2>+<B_{\rm r}^2>}
\end{eqnarray}
Its values are ranged between $-1$ (pure radial field) and $1$ (pure horizontal field).
Several authors did mention this parameter as a way to include the geometrical aspects of the field in order to implement its effect in a 1D code. Fig. \ref{Goldreich-Beta} shows the radial dependence of the anisotropy factor for the considered configuration: it can be immediatly understood that prescribing a realistic value of this parameter cannot be achieved without starting from a genuine latitudinally-dependent magnetic field.
\section{THE MODIFIED MECHANICAL BALANCE}
In this section, we first examine the physical quantities associated with the presence of the magnetic field that are likely to modify the mechanical balance. Their two-dimensional profiles are drawn to highlight their latitudinal dependence and their radial mean profiles, obtained by averaging over the latitudes, are discussed in the relevant cases. Then, the perturbations induced by the inclusion of the Lorentz force in the hydrostatic balance are derived and discussed.
\subsection{Physical quantities modifying the classical equilibrium} 
\subsubsection{The Magnetic Pressure}
\begin{figure}
\begin{center}
\includegraphics[width=0.475\textwidth]{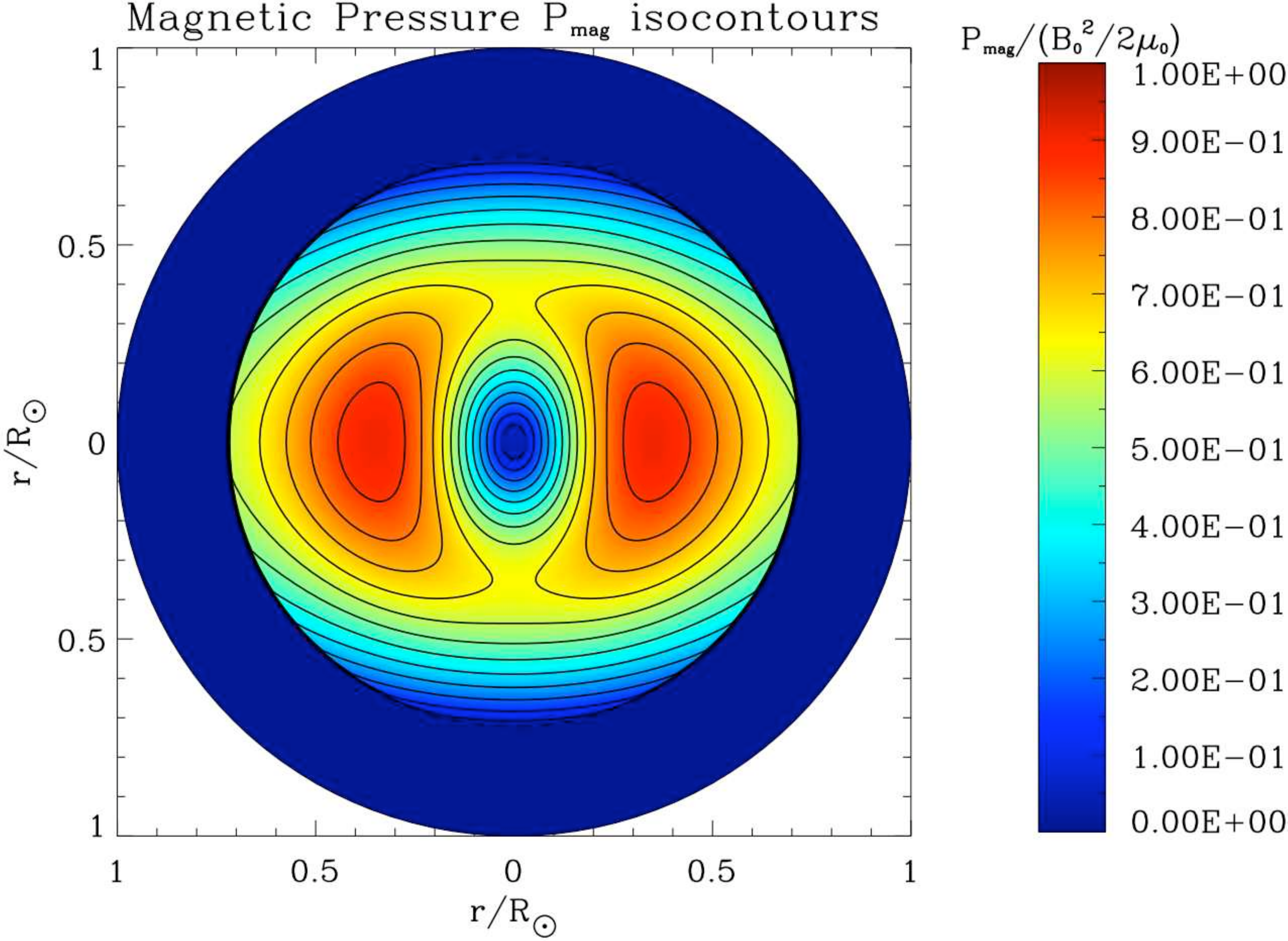}
\caption{Isocontours of the magnetic pressure $P_{\rm mag} (r, \theta)$ normalized to $B_0^2 / 2 \mu_0$ in meridional cut.
\label{Pmag}}
\end{center}
\end{figure}

The magnetic pressure is defined by
$P_{\rm{mag}} = \mbf{B}^2/{2 \mu_0}$ 
and its normalized profile with respect to $B_0^2/ 2 \mu_0$ is given in Fig. \ref{Pmag}. This is a quantity of special interest that may play a key role over secular time-scales. At first, it disturbs the hydrostatic balance through the contribution of its gradient to the Lorentz force.
From Fig. \ref{Pmag}, we can infer, owing to the direction of the magnetic pressure gradient, which is orthogonal to the magnetic pressure iso-surfaces, that the main effect of $P_{\rm mag}$ is to expel the gas from high magnetic pressure regions (red regions) and concentrate it in low magnetic pressure regions (blue regions).
The Sun remains in the strong $\beta$ regime ($\beta = P_{\rm{gas}}/P_{\rm{mag}}$), $\beta \geqslant 10^3$ in the radiative zone where the gas pressure strongly dominates the magnetic one. So, the Lorentz force is assumed to be only a perturbation compared with the self-gravitation of the star.
\begin{figure*}
\begin{center}
\includegraphics[width=0.475\textwidth]{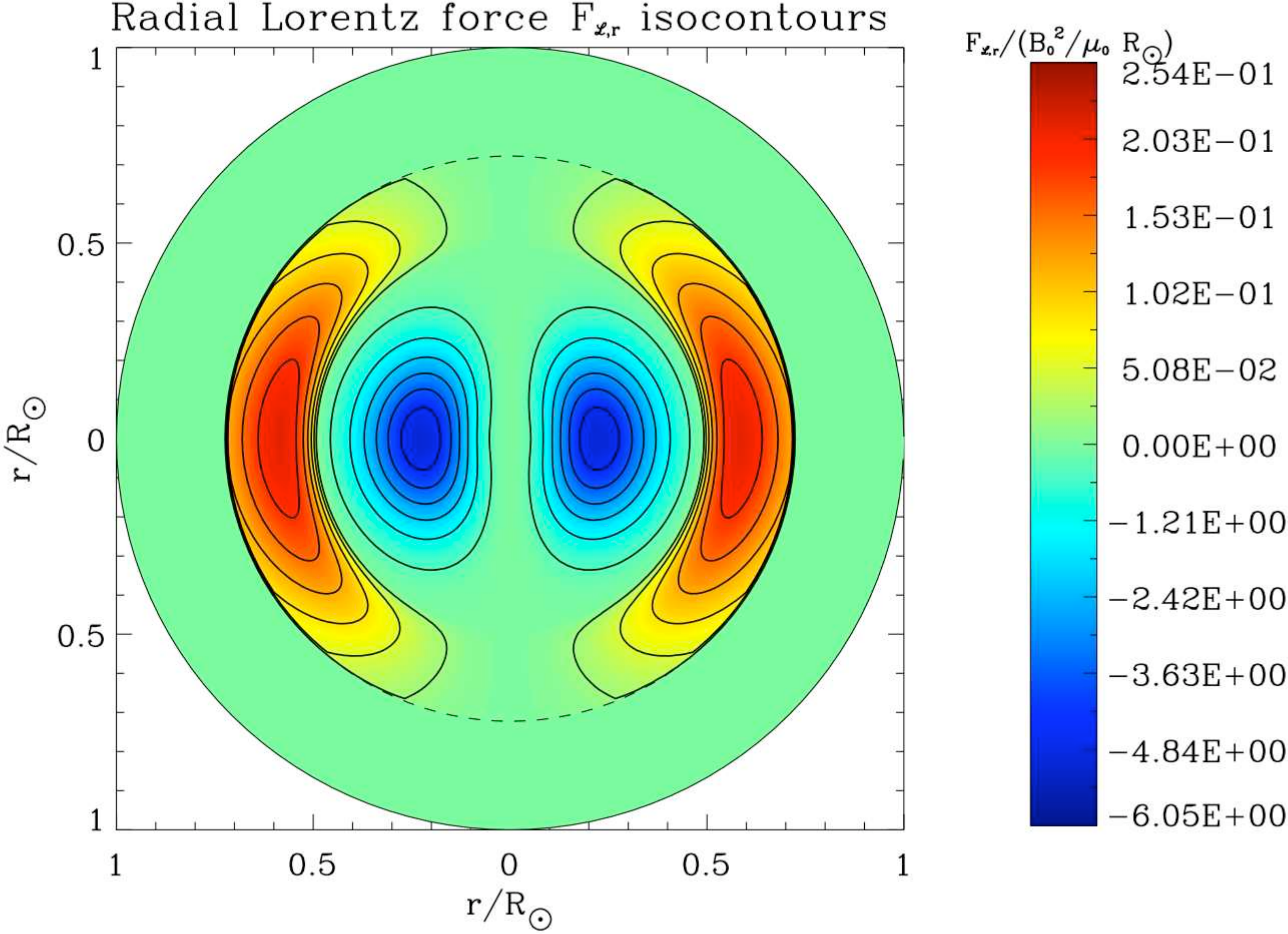}
\includegraphics[width=0.475\textwidth]{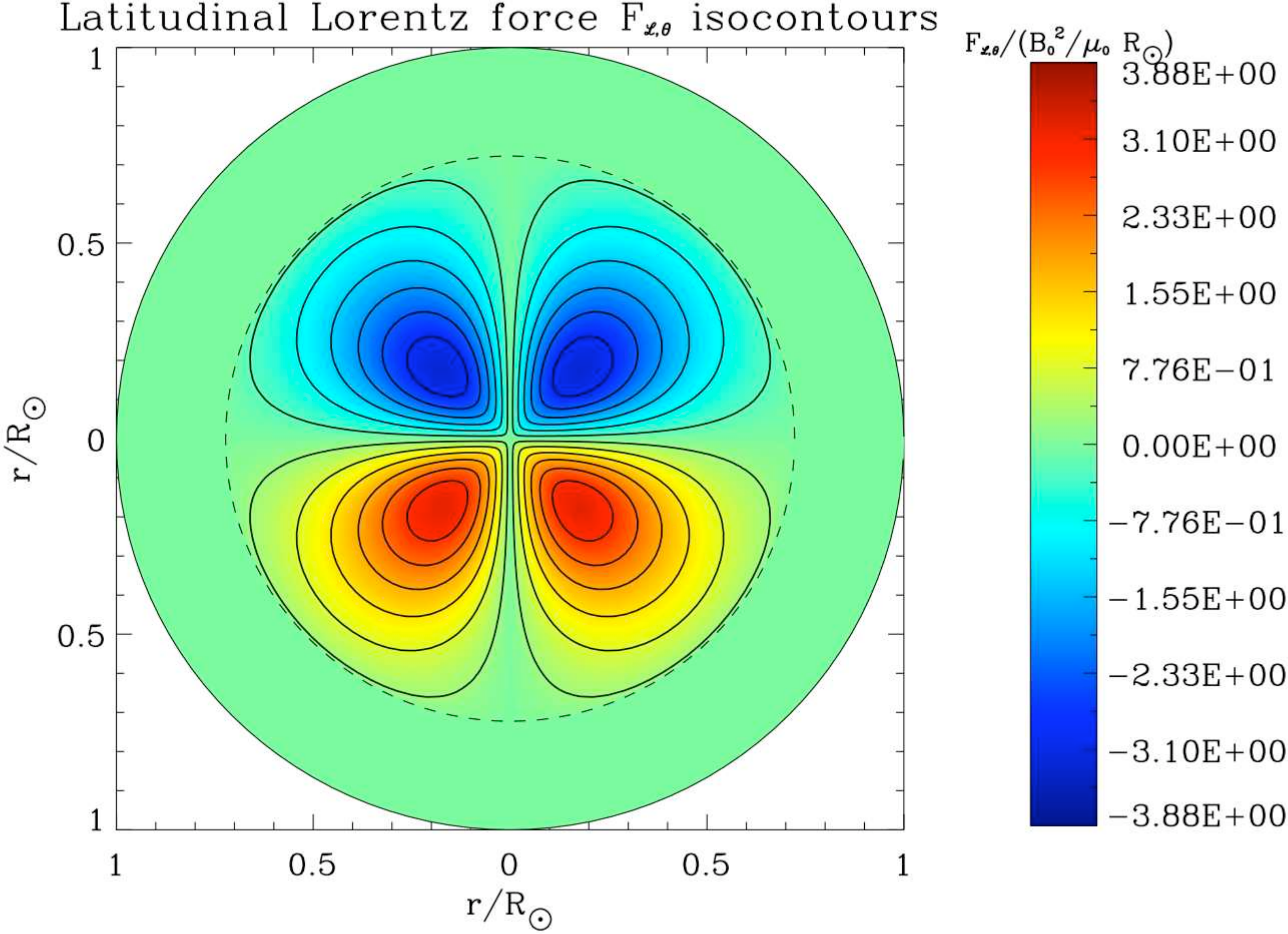}
\caption{a) Isocontours of the radial Lorentz force $\textit{\textbf{F}}_{\!\mathbf{{\mathcal{L}}};r} (r, \theta)$ normalized to $B_0^2 / \mu_0 R_\odot$ in meridional cut. b) Isocontours of the latitudinal Lorentz force $\textit{\textbf{F}}_{\!\mathbf{{\mathcal{L}}};\theta} (r, \theta)$ normalized to $B_0^2 / \mu_0 R_\odot$ in meridional cut.
\label{FL}}
\end{center}
\end{figure*}
\subsubsection{The Lorentz Force}
The Lorentz force is here axisymmetric. It modifies the stellar structure through the perturbation of the hydrostatic balance both in the radial and latitudinal directions. Its expression is given by:
\begin{eqnarray}
\mbf{F}_{\mathcal L} = - {\nabla (B^2}/{2 \mu_0}) + \frac{1}{\mu_0} \left(\mbf{B}\cdot \bnab \right) \mbf{B}
\end{eqnarray} 
To understand precisely its influence, its components along $\widehat{\mbf e}_r$ and $\widehat{\mbf e}_{\theta}$ normalized with respect to $B_0^2 / \mu_0 R_\odot $ are plotted in Fig. \ref{FL}. Furthermore, the averaged Lorentz force is drawn in Fig. \ref{MagTension}.
The Lorentz force exerts a {centripetal} influence in the inner part of the radiative zone (up to $0.47 \: R_\odot $) and a {centrifugal} one in the external part of the star. 
Furthermore, the latitudinal component of the Lorentz force is {negative in the northern hemisphere and positive in the southern one, then directed towards the poles. This will necessitate a counterbalancing gravitational force directed towards the equator, that will increase the density at low latitudes}. Therefore, the sphere deformation will be oblate. 
\subsubsection{Effect of the Magnetic Tension}
\begin{figure*}
\includegraphics[width=0.45\textwidth]{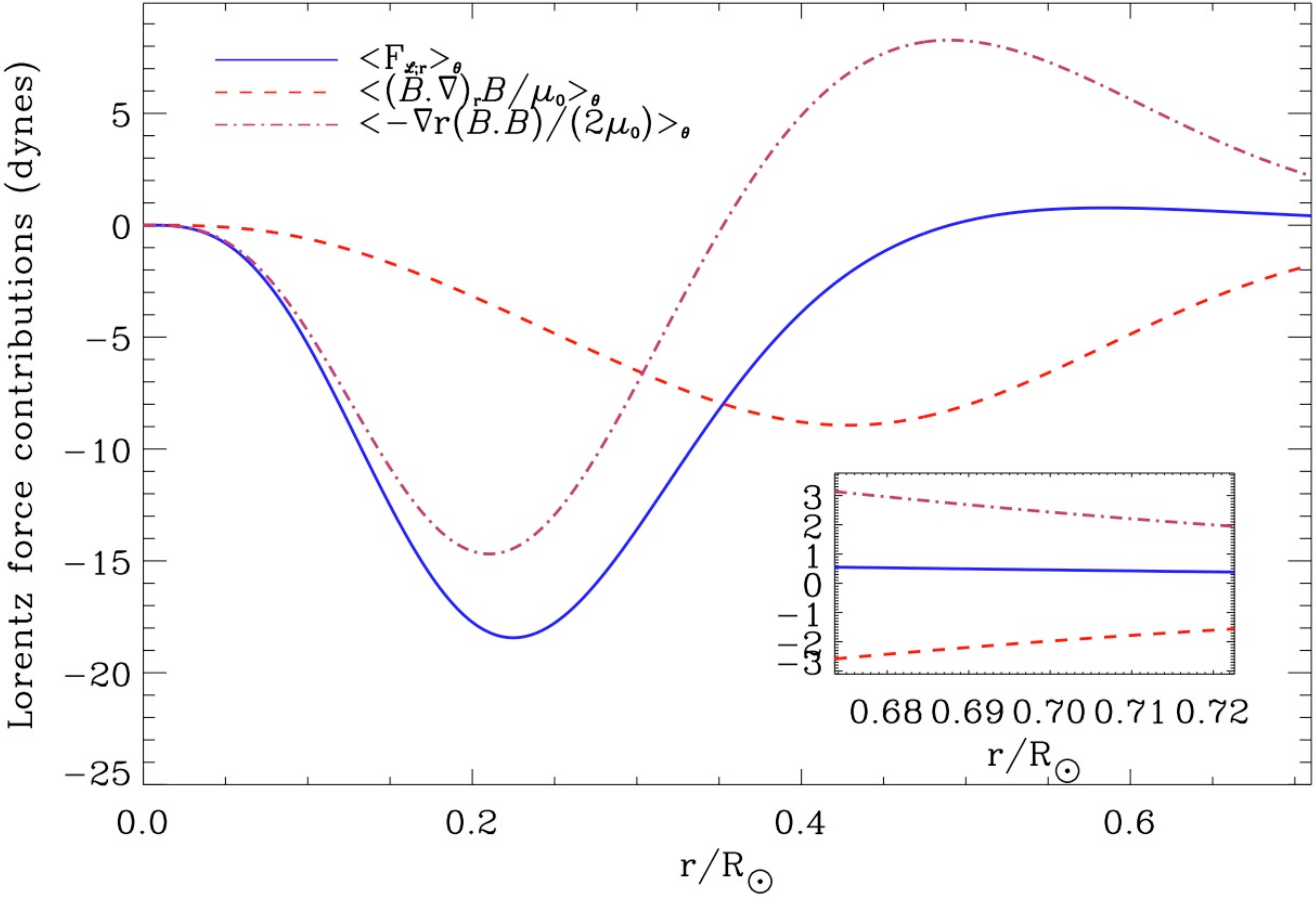}
\includegraphics[width=0.45\textwidth]{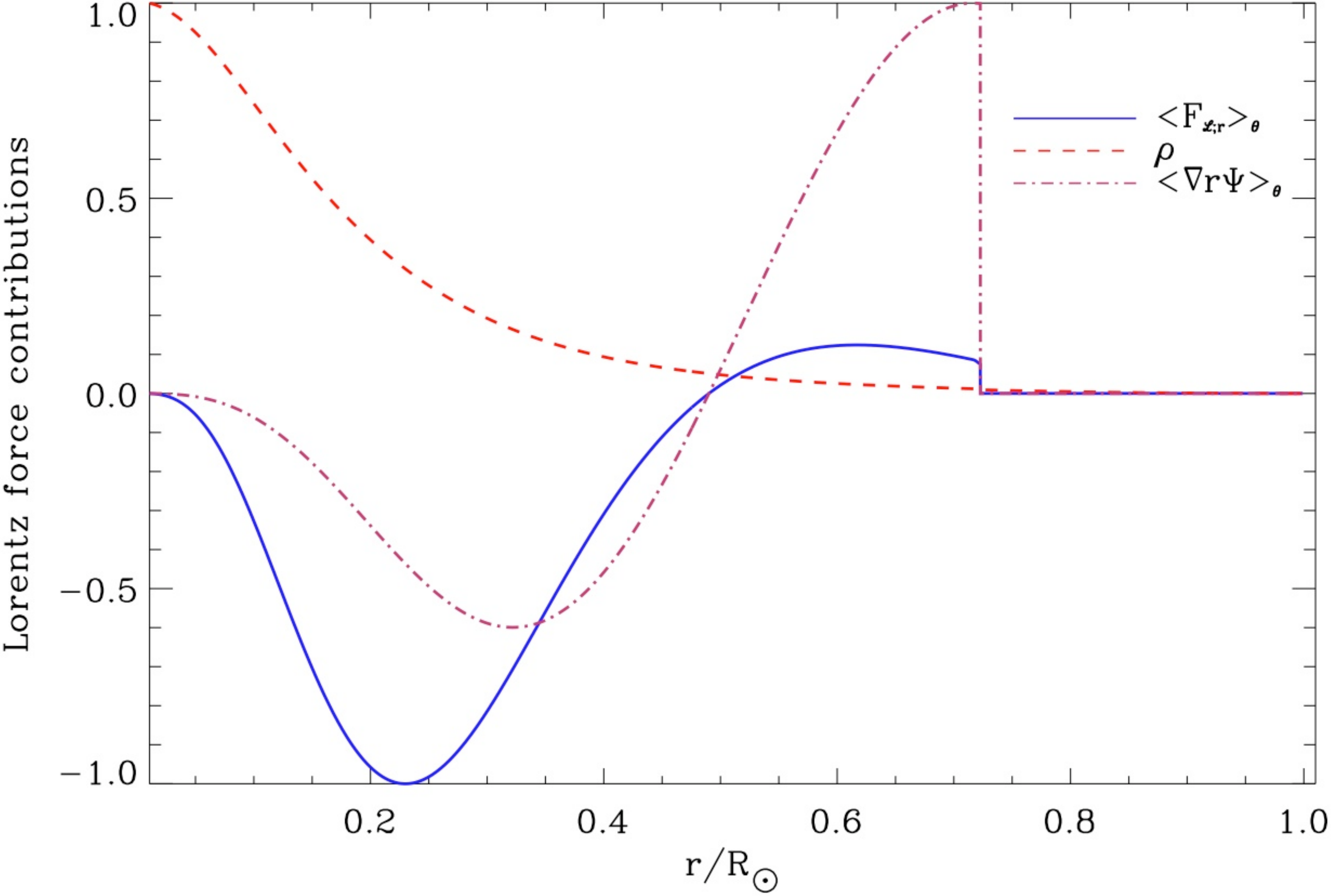}
\includegraphics[width=0.45\textwidth]{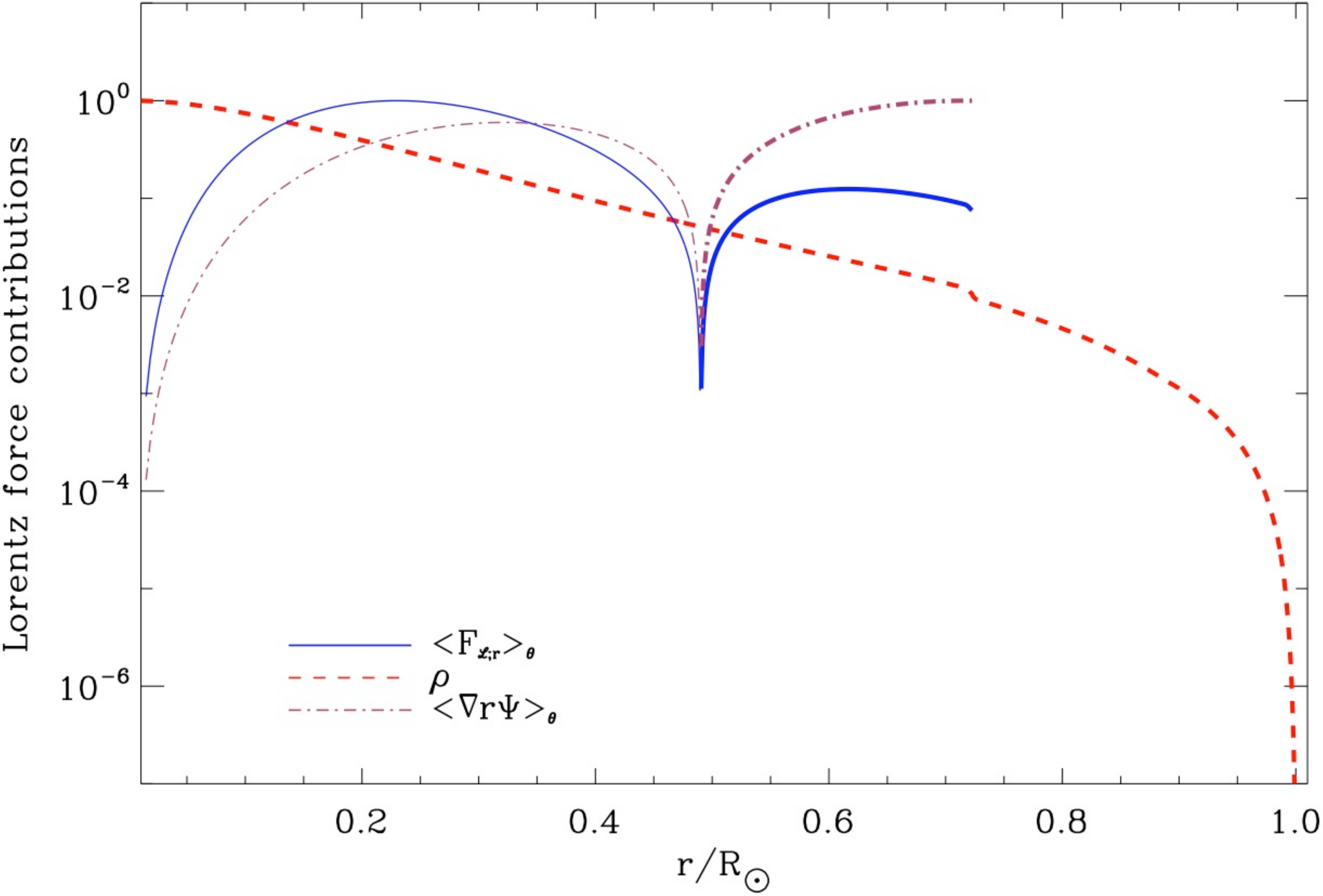}
\caption{a) Mean radial Lorentz force $\left< {F}_{\mathcal L; r} \right>_\theta$ (solid line), and its magnetic pressure gradient contribution compared with the magnetic tension term. b) Mean radial Lorentz force (solid line), and the contribution of the factor $\bar{\rho}$ and of $\left<\bnab_r \: \Psi \right>_\theta=\left<\partial_r\Psi\right>_\theta$ (normalized to their maximum). c) Same than the previous one in log scale. Bold lines represent positive values whereas thin lines represent negative ones.
\label{MagTension}}
\end{figure*}

We now focus on the second part of the Lorentz force, namely the magnetic tension one, which is defined by
\begin{eqnarray}
\mbf{F}_\mathcal{L}^T \equiv \frac{1}{\mu_0} \left(\mbf{B}\cdot \bnab \right) \mbf{B} = \mbf{F}_\mathcal{L} + \bnab P_{\rm{mag}}.
\end{eqnarray}
Therefore, a complete treatment of the mechanical balance modification by the Lorentz force cannot be achieved by taking into account only the magnetic pressure gradient. 
In Fig. \ref{MagTension}a, we draw the average over latitudes of the radial component of the complete Lorentz force together with the averaged magnetic pressure radial gradient and the averaged radial component of the magnetic tension force. 
We note that the magnetic pressure gradient dominates the magnetic tension in the internal part of the stars. 
However, the latter strength becomes of the same order of magnitude that those of the magnetic pressure term, in particular on the symmetry axis and in the vicinity of the surface where one counterbalancing each other as the total Lorentz force tends to vanish. 
In the model presented here 
the Lorentz force can be stated as 
\begin{eqnarray}
\mbf{F}_\mathcal{L} = \beta_0 \: \bar{\rho}\: \bnab \: \Psi
\end{eqnarray}
from which it arises that the Lorentz force vanishes if the density tends towards zero. The respective contributions of the two factors $\bar{\rho}$ and $\left<\bnab_r \: \Psi\right>_{\theta}=\left< \partial_r \:\Psi\right>_{\theta}$ to the radial component of the Lorentz force are drawn on Fig.  \ref{MagTension}b. In Fig. \ref{MagTension}c, it is clearly shown that the Lorentz force is mainly driven by the density evolution.
\subsection{Influence of the deep field on the Solar Structure}
\subsubsection{The perturbative approach}
The previous section has shown through the $\beta$-parameter that the Lorentz force is only a perturbation compared with the gravitational one and with the gaseous pressure gradient. So a first-order perturbative treatment can be performed to compute the structural perturbations associated with the modification of the hydrostatic equilibrium due to the magnetic field. We now derive the modified Poisson's equation for the perturbed gravific potential and deduce the density, pressure and radius related perturbations. 

This work can be applied to any star  in the high $\beta$ regime. In this case the stellar structure is only weakly two-dimensional and the horizontal variations of all quantities are small and smooth enough to allow their linearization.  Their generalized expressions are given in Appendix A. Let us recall that \cite{swe50} was the first to derive this result for a general perturbing force, \cite{mos74} having introduced the special case of the Lorentz force in the case of a poloidal field while the case of a general axisymmetric configuration (both poloidal \& toroidal) has been treated in Mathis \& Zahn (2005).
Note also that since the Lorentz force is proportional to $\left(\bnab \times \mbf{B} \right)\times \mbf{B}$, the perturbations amplitude are proportional to the square of the field's strength.

Though the quantities estimated in the Appendice A have been computed up to $l=8$ by \cite{aja06} using the theory of figures, our purpose is different. We prepare a complete  global MHD solar model and would like only here to estimate the relative influence of the different terms. We are interested in the dipolar mode of the magnetic field, so we  derive the $J_l$ for the modes $l=0$ and $l=2$ only, but the method could be extended for a more general angular geometry.

The Sweet's equation for the gravitational potential is integrated  using a finite difference, fourth-order Runge-Kutta scheme coupled to a shooting method to deal with an initial value problem. {The gravitational potential perturbation has been obtained by interpolating over an array with ten thousand  constant radial steps. It has then been interpolated back on the original \cesam mesh. This latter, more resolved in the subsurface, allows us to compute the perturbations of the other parameters with a greater accuracy, which was not necessary for the gravitational potential since that one does not vary steeply in this region (cf. Fig. 6).} Derivatives have been computed using a second order centered-difference scheme with a quadratic extrapolation outside the domain to conserve the precision at the boundaries. The shooting method is assumed to converge when the relative error between two consecutive models (of gravitational potential perturbation) is lower than $10^{-9}$. {It has been checked that the results are not sensitive to the number of interpolation points, nor to the relative error}.
\subsubsection{Results}
\begin{figure*}
\begin{center}
\includegraphics[width=0.475\textwidth]{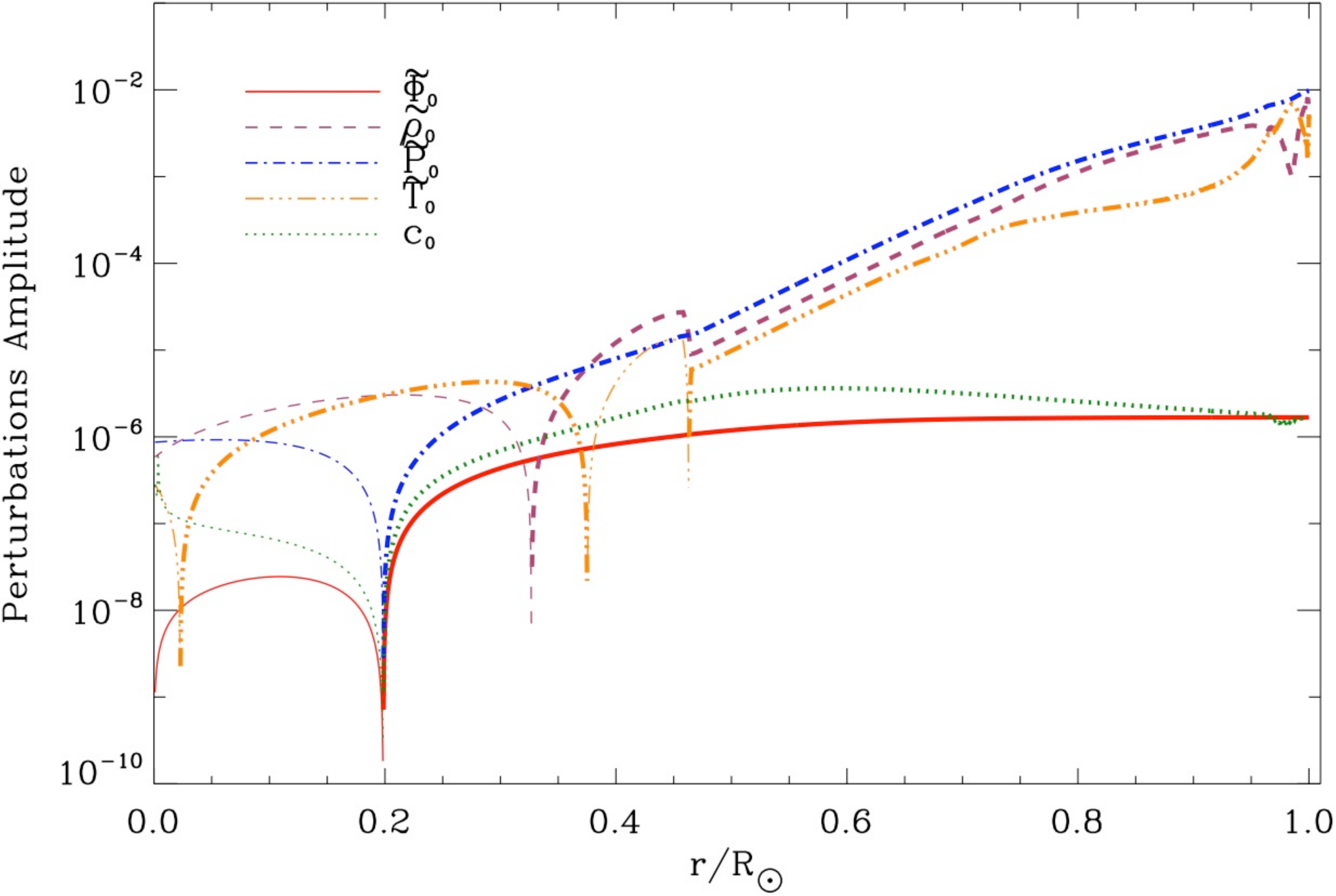}
\includegraphics[width=0.475\textwidth]{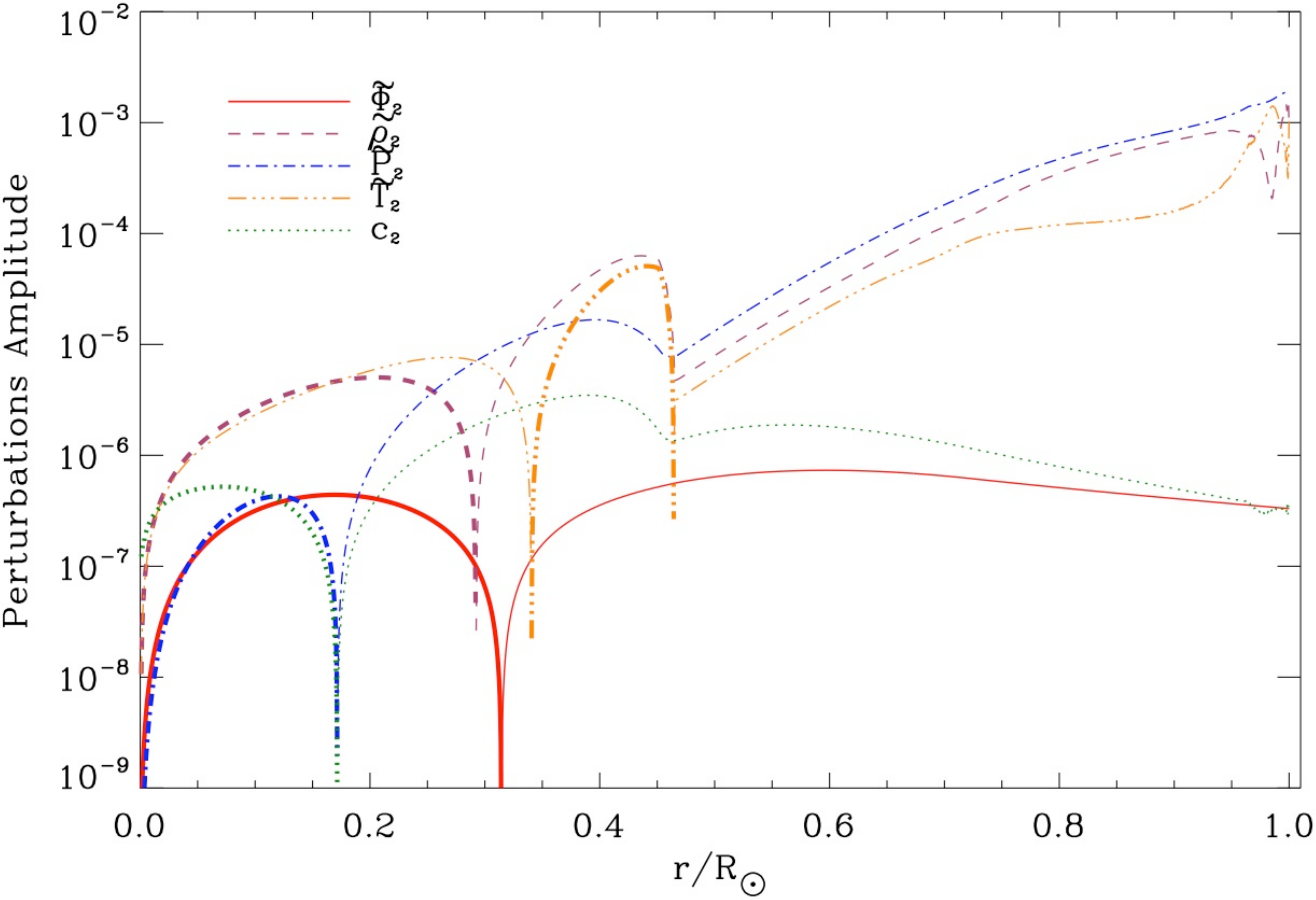}
\caption{Left panel: Normalized perturbations for the mode $l=0$ of gravitational potential, density, pressure, temperature and radius as a function of the radius. Right panel:  Normalized perturbations for the mode $l=2$ of gravitational potential, density, pressure, temperature and radius as a function of the radius. Negative values are drawn using thin curves while positive ones are drawn using bold curves. 
\label{flucts_l0}}
\end{center}
\end{figure*}

Let us define the normalized perturbation of a scalar $X$ to its unperturbed value by
\begin{eqnarray}
\tilde{X}_{l} = \widehat{X}_{l}/ X_{0}.
\end{eqnarray}
Results for the normalized perturbations in gravitational potential ($\tilde{\Phi}_{l}$), density ($\tilde{\rho}_{l}$), pressure ($\tilde{P}_{l}$), and radius ($c_{l}$) are shown in Fig. \ref{flucts_l0}  for the modes $l=0$ and $l=2$. {Bear in mind that the sign of the normalized perturbation in gravitational potential is the opposite of its unperturbed value, that one being negative.} 
Negative values are drawn using thin lines while positive ones are drawn using bold lines.\\

\begin{figure*}
\begin{center}
\includegraphics[width=0.475\textwidth]{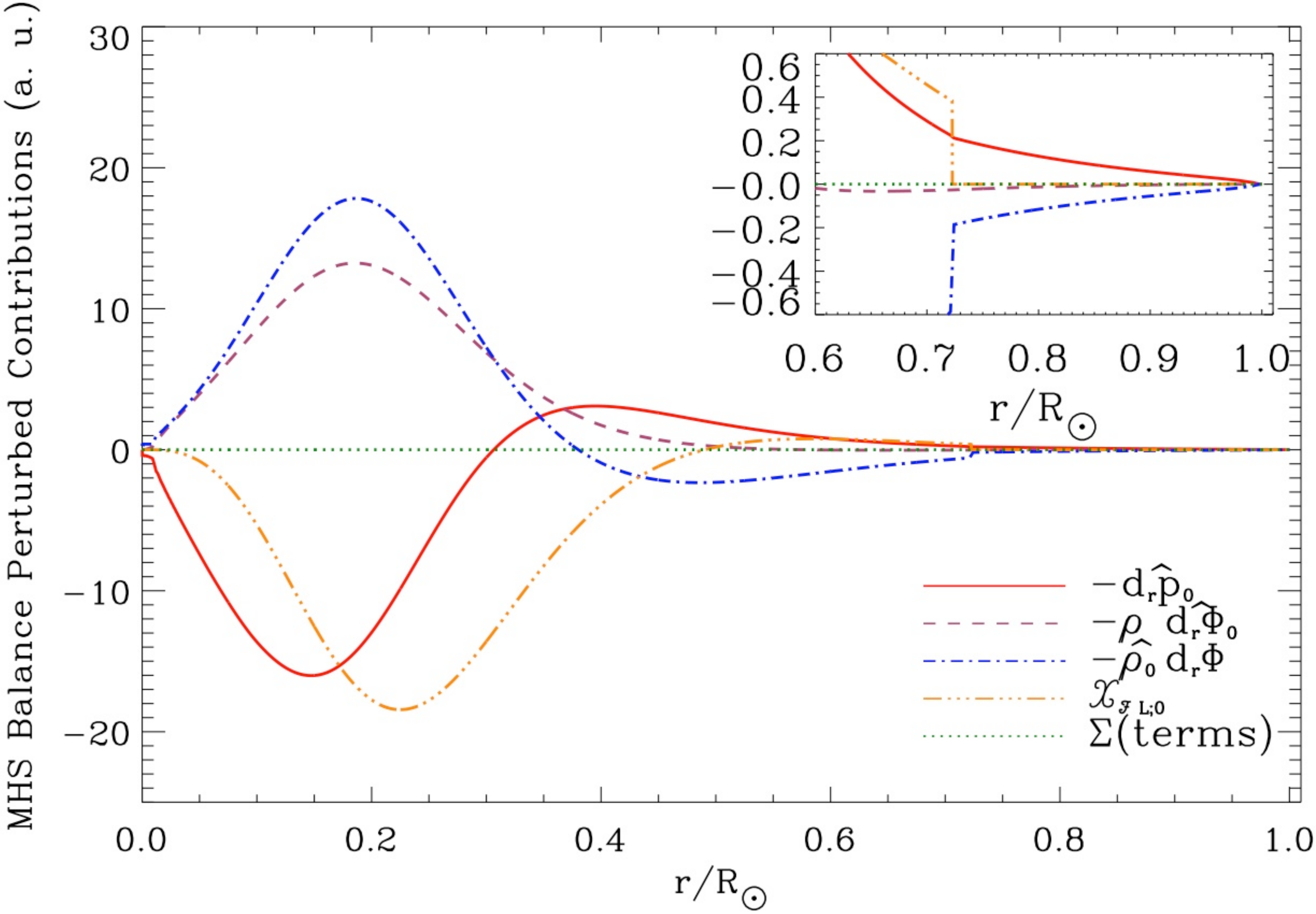}
\caption{Perturbations of the terms constituting the hydrostatic balance for the mode $l=0$, and their vanishing sum expressing the new equilibrium (here $\phi=-G M_r/r$).
\label{mhsbalance}}
\end{center}
\end{figure*}

The perturbations to the mean hydrostatic balance ($l=0$ terms) induced by the magnetic field are drawn on Fig. \ref{mhsbalance}. The sum of the different terms constituting the magneto-hydrostatic balance (Eq. \ref{Per1}) is also drawn as a verification of the new equilibrium settled in presence of magnetic field. One can observe the following features. First, the induced perturbation is of the same order of magnitude  than the cause of the perturbation, namely here the Lorentz force coefficient $\mathcal{X}_{\mbf{F}\!\!_{\mathcal{L}};0}$. {Second, the structure responds by increasing the perturbation in gravity (the term $\rho_0 \: \bnab \Phi^{(1)}$ being positive since $- \rho \: d \widehat{ \Phi_0}/dr <0$) where the Lorentz force has its major contribution (i. e. in the central part). The same remark can be drawn for the perturbation in density. On the contrary, the pressure force perturbation is opposed to the Lorentz force in this region. \\
The fact that the radial component of the Lorentz force changes sign leads to opposite effects in the outer part of the star. 
}
 \\
{Finally, the value of the quadrupolar gravitationnal multipolar moment $J_2$ is positive, which confirms our preceding intuition based on the sign of the Lorentz force, that the sphere deformation would be oblate. Furthermore, its value is $3.3 \times 10^{-7}$ for the 2 MG magnetic field imposed. In contrast, actual estimates for the gravitational multipolar moment due to the solar internal rotation are close to the value of $2.2 \times 10^{-7}$ (see e.g. \cite{Pat96, rox01}). Assuming that the values would be of the same order of magnitude for a young or the present Sun, with a magnetic field of same intensity, it thus arises that an internal magnetic field with MG strength might be a non-negligible source of oblateness.}

\section{THE MODIFIED ENERGETIC BALANCE}
We look first for the temperature perturbation induced by the presence of the magnetic field. Following \cite{kip90}, we introduce the general equation of state for the stellar plasma
\begin{equation}
\frac{{\rm d}\rho}{\rho}={\alpha_{s}}\frac{{\rm d}P}{P}-{\delta_{s}}\frac{{\rm d}T}{T}+{\varphi_{s}}\frac{{\rm d}\mu_{s}}{\mu_{s}},
\label{EqState}
\end{equation}
where $\alpha_{s}\!=\!\left(\partial\ln\rho/\partial\ln P\right)_{T,\mu_{s}}$, $\delta_{s}\!=\!-\left(\partial\ln\rho/\partial\ln T\right)_{P,\mu_{s}}$ and $\varphi_{s}\!=\!\left(\partial\ln\rho/\partial\ln\mu_{s}\right)_{P,T}$. 
In the framework in which the volumetric Lorentz force is a perturbation compared with the gravity, the stellar temperature ($T$) and the mean molecular weight ($\mu_{s}$) can be expanded like $\phi$, $\rho$, and $P$ following the expressions:  
\begin{eqnarray}
T\left(r,\theta\right) 
&=&T_{0}\left(r\right)+T^{\left(1\right)}\left(r,\theta\right)\nonumber \\ 
&=& T_{0}\left(r\right)+\sum_{l\ge0}{\widehat T}_{l}\left(r\right)P_{l}\left(\cos\theta\right);\quad\\
\mu_{s}\left(r,\theta\right) 
&=& \mu_{s;0}\left(r\right)+\mu_{s}^{\left(1\right)}\left(r,\theta\right)\nonumber \\ 
&=& {\mu}_{s;0}\left(r\right)+\sum_{l\ge0}{\widehat \mu}_{s;l}\left(r\right)P_{l}\left(\cos\theta\right)\!.\;\quad
\end{eqnarray}
Using the linearization of Eq. (\ref{EqState}) around the non-magnetic state, we then get for each $l$:
\begin{equation}
\frac{{\widehat \rho}_{l}}{\rho_{0}}=\alpha_{s}\frac{\widehat P_{l}}{P_{0}}-\delta_{s}\frac{{\widehat T}_{l}}{T_{0}}+{\varphi_{s}}\frac{{\widehat \mu}_{s;l}}{\mu_{s;0}}
\end{equation}
that leads finally to
\begin{equation}
{\widehat T}_{l}=\frac{T_{0}}{\delta_{s}}\left[\alpha_{s}\frac{{\widehat P}_{l}}{P_{0}}-\frac{{\widehat\rho}_{l}}{\rho_{0}}+{\varphi}_{s}\frac{{\widehat \mu}_{s;l}}{\mu_{s;0}}\right].
\end{equation}
${\widetilde T}_{0}={\widehat T}_{0}/T_{0}$ and ${\widetilde T}_{2}={\widehat T}_{2}/T_{0}$ are respectively plotted in Figs. \ref{flucts_l0}  for the considered solar model. We assume that $\alpha_{s}=\delta_{s}=1$ which is acceptable for main-sequence stars and we choose here not to take into account the mean molecular weight fluctuation ($\varphi_s=0$) (see Kippenhahn \& Weigert 1990). 
\subsection{Physical Quantities Modifying The Energetic Balance}
\begin{figure*}
\begin{center}
\includegraphics[width=0.45\textwidth]{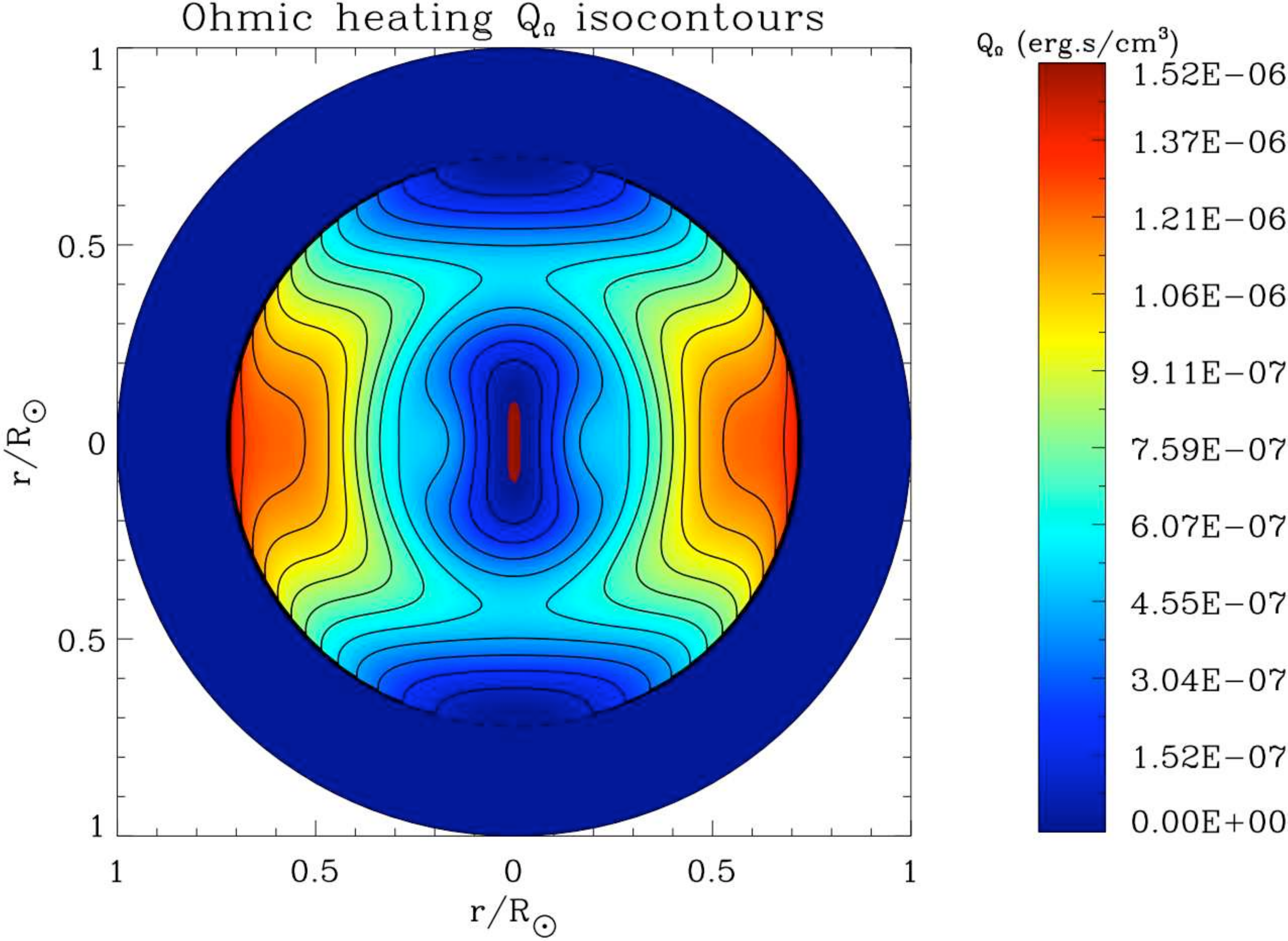}
\includegraphics[width=0.45\textwidth]{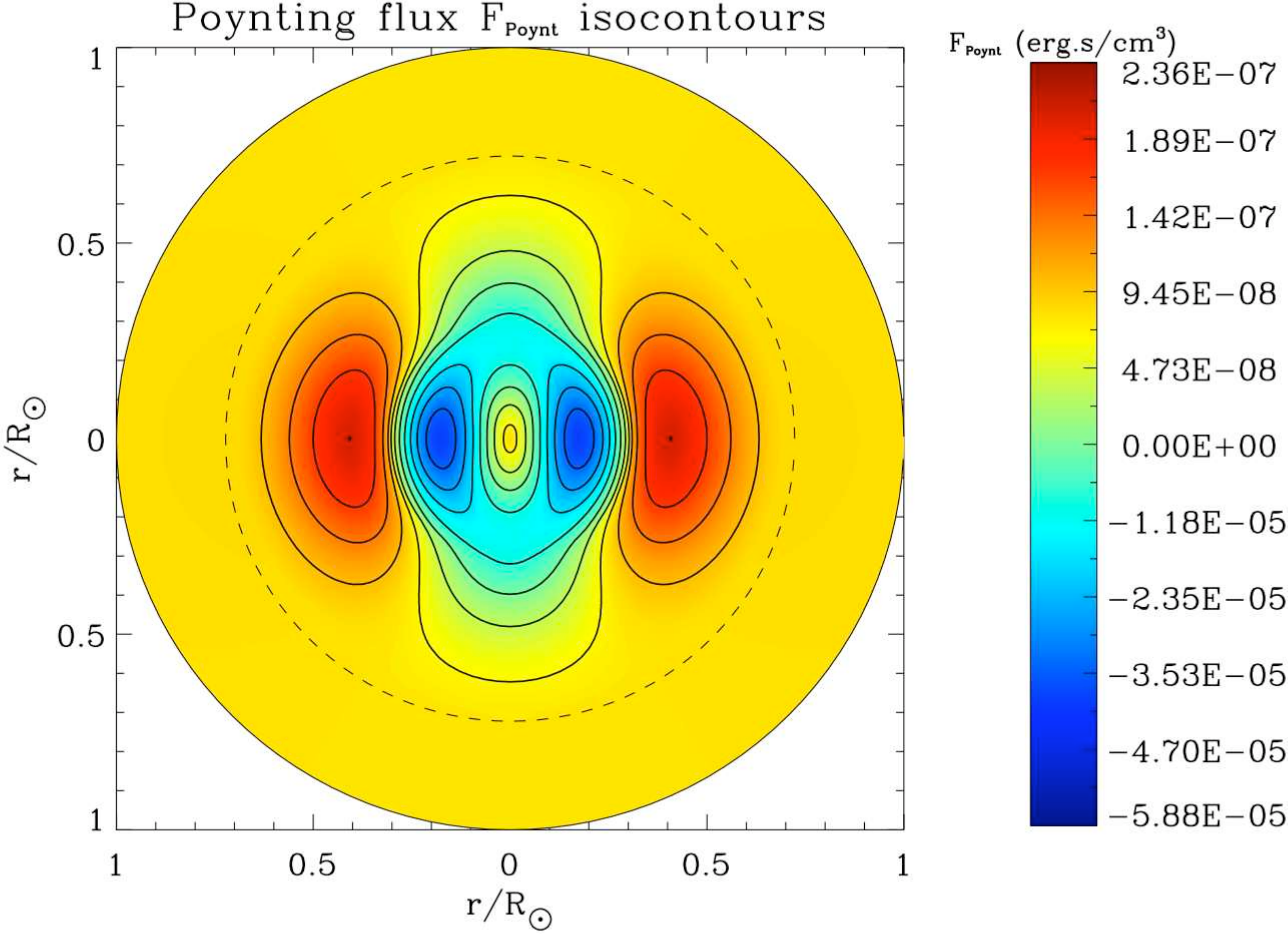}
\caption{a) Isocontours of the Ohmic heating $Q_{{\rm Ohm}} (r, \theta)$ in meridional cut. b) Isocontours of the Poynting's flux $F_{\rm{Poynt}} (r, \theta)$ in meridional cut.
\label{QOhm}}
\end{center}
\end{figure*}

The two terms that have now to be examined are respectively the volumetric ohmic heating and the flux transported by the electromagnetic field, namely the Poynting's flux.
The Ohmic heating becomes in the case of an isotropic magnetic diffusivity ($\eta$)
\begin{eqnarray}
Q_{{\rm Ohm}} (r, \theta) = \mu_0\: \eta \: \mbf{j}^2\left(r,\theta\right).
\end{eqnarray}
Considering the following temperature-dependent law for $\eta$ \citep{spi62}
\begin{eqnarray}
\eta = 5.2 \times 10^{11}\: \log \Lambda \:T^{-3/2} \;\rm{cm}^{2} \rm{s}^{-1}
\label{magdiffusiv}
\end{eqnarray}
where we take for the coulombian logarithm $ \log \Lambda \approx 10$,
we can compute semi-analytically the Ohmic heating (see Fig. \ref{QOhm}a). 
The Ohmic heating increases steadily as we go up to the radiation-convection transition.
The Poynting's flux  is given by
$F_{\rm{Poynt}}=\bnab \cdot \mbf{S}$,
{where the Poynting's vector is defined by
$\mbf{S} =  \mbf{E} \times \mbf{B}/ \mu_0$.}
In the static case, the simplified Ohm's law gives 
$\mbf{j} = \sigma \mbf{E}$.
Using the identity between the conductivity ($\sigma$) and the magnetic diffusivity $ \eta = \left(\mu_0 \sigma \right)^{-1}$, 
the Poynting's flux reduces then to 
\begin{eqnarray}
F_{\rm{Poynt}} =  \bnab\cdot\left(\eta \:\FLor \right).
\end{eqnarray}
This term is shown for the considered configuration in Fig. \ref{QOhm}b. 
\begin{figure*}
\begin{center}
\includegraphics[width=0.45\textwidth]{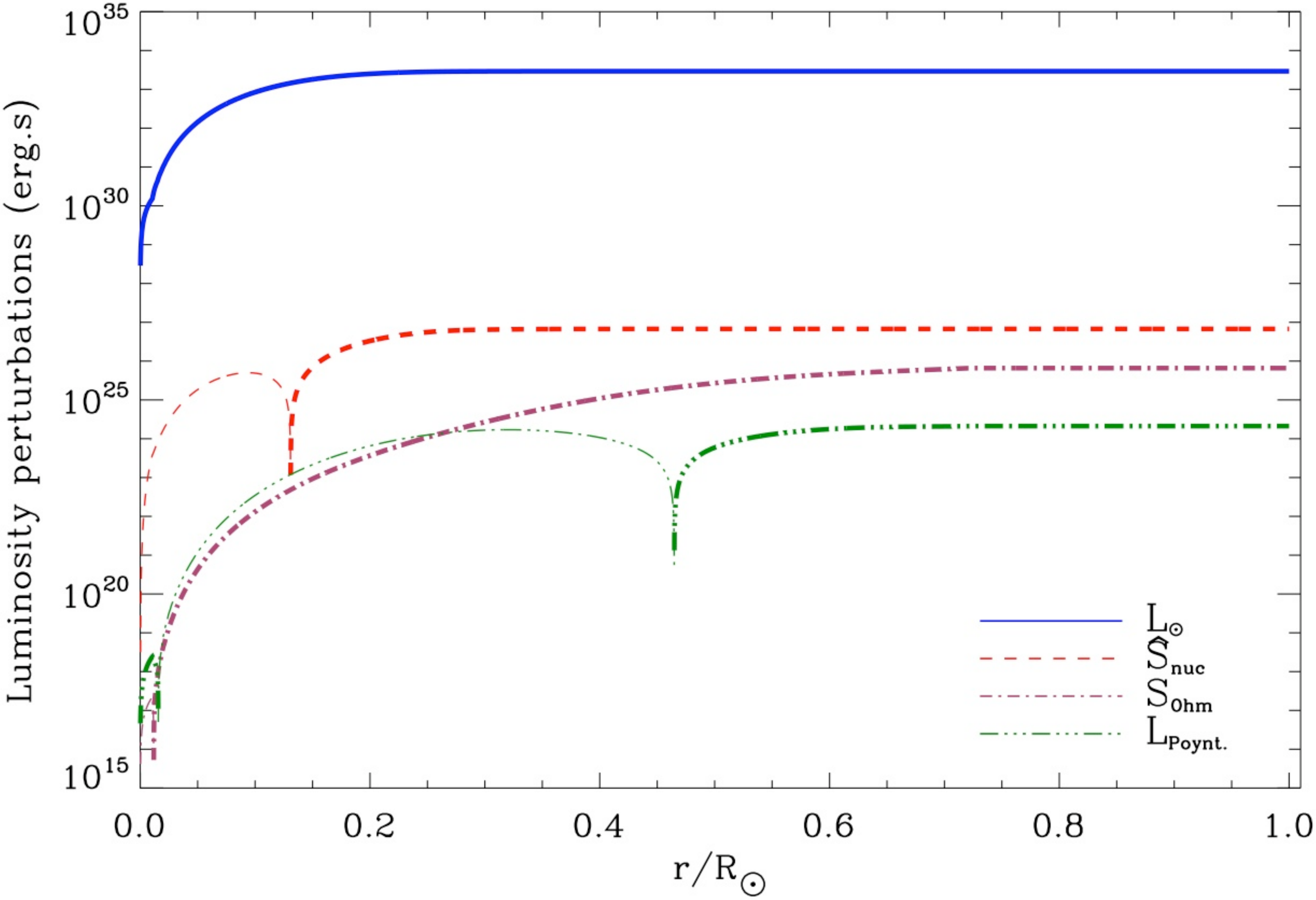}
\caption{Contribution to the luminosity perturbation through Ohmic heating (dash-dotted line), Poynting's flux (dash-dot-dotted line) and nuclear reaction rates perturbations (dashed line) as a function of the radius. Negative values are drawn using thin curves while positive ones are drawn using bold curves.
\label{luminmag}}
\end{center}
\end{figure*}
\subsection{Perturbation of the Energetic Balance}
Let us now consider the modification of the energetic balance. First, according to Poynting's theorem:
\begin{eqnarray}
\PA{U_{\rm{mag}} }{t} = -Ê\bnab\cdot \mbf{S} - Q_{{\rm Ohm}},
\end {eqnarray}
where $U_{\rm{mag}}$ is the electromagnetic energy density {and $\mbf{S}$ the Poynting's vector}. Then, the classical energy equation
\begin{eqnarray}
\PA{L}{M_r} &=& \varepsilon - \frac{1}{\rho} \PA{U_{\rm{int}}}{t} +\frac{P}{\rho^2} \PA{\rho}{t}
\end{eqnarray}
($L$ being the total luminosity, $M_r$ the mass, $\varepsilon$ the specific energy production rate per unit mass, and $U_{\rm{int}}$ the gas internal energy density) is modified into\footnote{In the case where the magnetic field is taken into account, the star loses its spherical geometry. Conversely, the luminosity ($L$) is defined as $L\left(r\right)=\int_{\Sigma}{\bf F}\cdot{\rm d}{\bf \Sigma}$ where ${\bf F}=-\chi{\bf \nabla}T$, $\chi$ being the thermal diffusivity. Therefore, we have to take the horizontal average of the right-hand side of the energy equation: 
${\bf\nabla}\cdot{\bf F} = \rho\varepsilon+{\bf\nabla}\cdot{\bf S}+Q_{{\rm Ohm}}$
to form Eq. (\ref{Lini}).}:
 \begin{eqnarray}
\PA{L}{M_r} &=& \left<\varepsilon - \frac{1}{\rho}\PA{U_{\rm{int}}}{t} -\frac{1}{\rho} \PA{U_{\rm{mag}} }{t} +\frac{P}{\rho^2} \PA{\rho}{t}\right>_{\theta}\nonumber\\
   		&=&  \left<\varepsilon -\frac{1}{\rho} \PA{U_{\rm{int}}}{t} +\frac{P}{\rho^2} \PA{\rho}{t} + \frac{1}{\rho}\bnab\cdot \mbf{S} +\frac{1}{\rho} Q_{{\rm Ohm}}\right>_{\theta}\!\!\!\!.\;\;\;\quad
		\label{Lini}
\end{eqnarray}
The purpose is here to quantify the exceeding amount of energy radiated away through the Poynting's flux {(this quantity representing the energy generated by the work of the Lorentz force which is converted in electromagnetic radiation)}\footnote{{In the static case, the Poynting's flux can be expressed as $\mbf{S} = \eta\: \mbf{j} \times \mbf{B} = \eta \:\mbf{F}_{\mathcal{L}}$.}} and dissipated by Ohmic heating.
We integrate the Ohmic heating and the Poynting's flux over the spherical shells of radius $r$
\begin{eqnarray}
{\mathcal S}_{{\rm Ohm}} (r) &=& \int_{0}^{r}\!\! \int_{\Omega}  Q_{{\rm Ohm}}(r', \theta')\, {\rm d}\Omega\,r'^2 {\rm d}r';\\
L_{\rm{Poynt}} (r) &=&\int_{0}^{r} \!\!\int_{\Omega} F_{\rm{Poynt}}(r', \theta')\, {\rm d}\Omega\,r'^2 {\rm d}r';\qquad
\end{eqnarray}
where ${\rm d}\Omega = \sin \theta' d\theta' d\phi' $, $r'$ thus ranging from $0$ to $r$, $\theta'$ from $0$ to $\pi$ and $\phi'$ from $0$ to $2 \pi$. 
This then allows us to compare their respective contribution on the total luminosity in function of the radius.

To get a complete diagnosis, we finally consider the modification of the specific energy production rate per unit mass ($\varepsilon$), which depends on $\rho$ and $T$, due to magnetic field.
First, the logarithmic derivative of $\varepsilon$ is expanded like the one of $\rho$ (cf. Eq. \ref{EqState} and Mathis \& Zahn 2004 and references therein):
\begin{equation}
\frac{{\rm d}\varepsilon}{\varepsilon}=\lambda\frac{{\rm d}\rho}{\rho}+\nu\frac{{\rm d}T}{T},
\label{EpsilonDef}
\end{equation}
where $\lambda=\left(\partial\ln\varepsilon/\partial\ln\rho\right)_{T}$ and $\nu=\left(\partial\ln\varepsilon/\partial\ln T\right)_{\rho}$. Then, like $\rho$ and $T$, we expand $\varepsilon$ as
\begin{eqnarray}
\varepsilon\left(r,\theta\right)
&=& {\varepsilon}_{0}\left(r\right)+{\varepsilon}^{\left(1\right)}\left(r,\theta\right) \nonumber\\
&=& {\varepsilon}_{0}\left(r\right)+\sum_{l\ge0}{\widehat{\varepsilon}}_{l}\left(r\right)P_{l}\left(\cos\theta\right).\qquad
\end{eqnarray}
Linearizing Eq. (\ref{EpsilonDef}) around the non-magnetic state, we obtain:
\begin{equation}
{\widehat\varepsilon}_{l}=\varepsilon_{0}\left[\lambda\frac{{\widehat \rho}_{l}}{\rho_0}+\nu\frac{{\widehat T}_{l}}{T_0}\right].
\end{equation}
Since $L_{{\rm Ohm}}+L_{\rm Poynt}\!<\!\!<\!L_{0}$, $L_{0}$ being the luminosity in the non-magnetic case  (here $L_{\odot}$ given in Fig. \ref{luminmag}), a perturbative approach can be used and we can expand the luminosity as
$L=L_{0}+{\widehat L}$,
where ${\widehat L}$ is the luminosity perturbation due to the magnetic terms.\\

Linearizing Eq. (\ref{Lini}) around the non-magnetic state, we get:
\begin{equation}
\frac{{\partial}L_{0}}{\partial M_r}=\varepsilon_{0}\,;
\label{L0}
\end{equation}
\begin{equation}
\frac{{\partial}{\widehat L}}{\partial M_r}=\left<{\varepsilon^{\left(1\right)}}+\frac{1}{\rho_{0}}{\bf\nabla}\cdot{\bf S}+\frac{1}{\rho_{0}}Q_{{\rm Ohm}}\right>_{\theta}.
\label{L1}
\end{equation}
Next, integrating Eq. (\ref{L1}), we obtain:
\begin{equation}
{\widehat L}\left(r\right)=\widehat{\mathcal S}_{\rm nuc}\left(r\right)+L_{\rm Poynt}\left(r\right)+{\mathcal S}_{{\rm Ohm}}\left(r\right),
\end{equation}
where
\begin{eqnarray}
\widehat{\mathcal S}_{\rm nuc}\left(r\right)
&=&\int_{\Omega}\varepsilon^{\left(1\right)}\rho_{0}{\rm d}\Omega
=\int_{0}^{m\left(r\right)}\left<\varepsilon^{\left(1\right)}\right>_{\theta}{\rm d}M_r\nonumber\\
&=&4\pi\int_{0}^{r}\left\{\varepsilon_{0}\left[\lambda\frac{{\widehat \rho}_{0}}{\rho_0}+\nu\frac{{\widehat T}_{0}}{T_0}\right]\right\}{\rho}_{0}\,{r'}^2{\rm d}r'.\;\;
\end{eqnarray}
Results obtained are given in Fig. \ref{luminmag}. The different contributions stay extremely small. Note that the solar case is very interesting because it demonstrates (if it is confirmed by a complete solar model) that these perturbations cannot affect the cyclic 11 year magnetic effect connected to the convective zone. Nevertheless, this luminosity effect is greater than the natural evolution of the solar luminosity ($10^{-8}$ on 100 years) and influences the long trend evolution. Moreover, this work shows the relationship between the luminosity and the radius fluctuations and could be also applied to a magnetic perturbation in the subsurface layers (see below).
\section{Surface Perturbations}
\subsection{Structural Surface Perturbations}
Using the continuity of the gravific potential ($\phi$) at the surface, we derive the expression to evaluate the gravitational multipolar moments $J_{l}$ (cf. Eq. \ref{Multipole}):
\begin{equation}
J_{l}=\left(\frac{R_\odot}{GM_\odot}\right)\widehat{\phi}_{l}\left(r=R_\odot\right).
\end{equation}
The radius perturbation can then be expressed as:
\begin{equation}
c_{l}\left(R_\odot\right)=\frac{\rho_{0}}{{{\rm d} P_{0}/{\rm d}r}}\left(\frac{1}{r}\frac{G M_\odot}{R_\odot}J_{l}+\frac{\mathcal{Y}_{\mbf{F}\!\!_{\mathcal L};l}}{\rho_{0}}\right).
\end{equation}
Their surface values, together with the perturbations in density, pressure and temperature are presented in Table \ref{tab1} for the mode $l=0$ and
for the mode $l=2$, for  the magnetic field buried in the solar radiative zone.
\begin{table}
\begin{center}
\caption{First order (FO) and second order (SO) structural perturbations due to the presence of a fossil field. Normalized modal surface perturbations for the gravitational potential (J), density ($\rho$), pressure (P), temperature (T), and radius (c).
\label{tab1}}
\begin{tabular}{lcccc}
\hline
\hline
FO Perturbation   &  Young Sun & SO Perturbation & Young Sun \\
\hline
 $J_{0} $& $ -1.68 \times10^{-6}$ &  $J_{2} $& $ 3.31 \times10^{-7}$ 
\\\hline
 $\tilde{\rho}_0 $& $4.57 \times10^{-3}$ &    $\tilde{\rho}_2 $& $-9.04\times 10^{-4}$
 \\\hline 
 $\tilde{P}_0 $& $ 9.78 \times10^{-3}$ &  $\tilde{P}_2 $& $-1.93 \times10^{-3}$
 \\\hline
 $\tilde{T}_0 $&$ 5.21 \times10^{-3}$ &  $\tilde{T}_2$&$ - 1.03 \times 10^{-3} $ 
  \\\hline
 $c_0 $&$ 1.73 \times10^{-6}$ &  $c_2 $&$ -3.42 \times10^{-7}$
 \\\hline
 \end{tabular}
 \end{center}
\end{table}
The effective temperature change owing to the presence of a large-scale magnetic field is {positive} in this solar case, of course the real effect for the present Sun could be different due to the total luminosity constraint. The same kind of study will be conducted for the present Sun after implementation in a stellar evolution code.
\subsection{Luminosity Perturbations}
The surface values of the respective contributions to the total luminosity are given in Table \ref{tab:luminmags}.
It is clear that the impact of the magnetic terms on the energetic balance appears to be very weak as it is well known.
\begin{table}
\begin{center}
\caption{Luminosity perturbations compared with the total present solar luminosity (reduced by 24\% at 500 Myr). Contributions of Ohmic heating $\left({\mathcal S}_{\rm Ohm} \right)$ and of Poynting's flux $\left( L_{\rm Poynt} \right)$ to the luminosity perturbation (in erg.s); contribution of the nuclear efficiency modification induced by the perturbation of the hydrostatic equilibrium (${\mathcal S}_{\rm nuc}$).
\label{tab:luminmags}}
\begin{tabular}{clr}
\hline \hline
Perturbation to the luminosity &  Young Sun  \\
\hline
$L_{\rm Ohm} $  		& $ 6.59 \times10^{25}   $
\\\hline
$L_{\rm Poynt}$    		& $ 2.12 \times10^{24}  $
\\\hline
$\widehat L_{\rm nuc}$     & $ 6.73 \times10^{26}  $
\\\hline
$L_{\rm tot}     $			& $ 2.93 \times10^{33}  $
\\
\hline
\end{tabular}
\end{center}
\end{table}
In this application, $L_{\odot}/ {\widehat L_{\rm nuc}}$ is about $10^{6}$, whereas ${\widehat L_{\rm nuc}} / \left(L_{{\rm Ohm}}+L_{\rm Poynt} \right)$ is around 40. 
Hence,  it is found that the direct contribution of the magnetic field to the change in the energetic balance through Ohmic heating or Poynting's flux is weak compared with the indirect modification to the energetic balance induced by the change in temperature and density over the nuclear reaction rate.
So as a first step, we can thus consider the impact of the large-scale magnetic field only upon the mechanical balance while the modification of the energetic one by the Ohmic heating and the Poynting flux is a higher order perturbation.
%
%
\section{CONCLUSION AND PERSPECTIVES}
In this paper, we do some progress in comparison with previous works which did not consider the topology of the magnetic field in stellar models. We show its consequences  and quantify the different terms appearing in the modified structural equations  of stellar evolution. To be quantitative, we have discussed  a possible configuration of fossil field located in the solar radiative zone. We have chosen this configuration for a young Sun supposing that  a transition region between convection and radiation like a tachocline is already established  (\cite{Gough09} and references therein). 

This paper clearly shows that one needs to take into account, in addition to the magnetic pressure both effects of magnetic pressure and magnetic tension in the hydrostatic balance through the Lorentz force. Fig. \ref{MagTension} shows the relative importance of each of them. In the vicinity of the base of the convective zone, the magnetic tension has an important role since it compensates the magnetic pressure gradient and tends to a force-free state; this is also the case near the magnetic field axis.\\
A first order-perturbative treatment has been performed. It puts in evidence the radial and latitudinal perturbations of the  structural quantities. This approach is valid in high-$\beta$ regime which is totally justified for the present case. The modal fluctuations in structural quantities, namely the gravific potential, the density, the pressure, the temperature, the gravitational multipole moments and the isobar radius are computed. 

Finally we show that  the Poynting's flux and the Ohmic heating have approximately the same order of magnitude. Nevertheless, their contributions to the modification of the energetic balance remain extremely weak in comparison with the modified nuclear energetic contribution. This last one is also small but may influence the determination of the present solar luminosity. 

This study is the first step to get a real MHD approach in stellar evolution. It allows to introduce the useful terms in a stellar code.  We will then consider the secular transport equations along the evolution  to estimate the real role of a magnetic configuration on the transport of angular momentum along the different stages of evolution. Such work will be done in a coming paper to estimate the magnetic field impact on the solar internal rotation profile and also to better quantify all the superficial observed quantities which will be measured by the coming space solar missions \textsc{SDO} and \textsc{Picard}. 

As it has already be emphasized, the formalism derived here to quantify the mechanical and energetic balance modifications is general and can also be applied to non-axisymmetric or/and time-dependent magnetic fields. This will be done in the future. Moreover the topology of the field and its intensity is only one example of a reasonable configuration, nobody knows today what it must be, so in the future certainly other configurations will be looked for. 

One can see already that the present order of magnitude of the different terms calculated for this young Sun in table 1 and 2 are not so far from what could be already estimated (see Rozelot 2009) but of course it stays a long route before introducing all the dynamical effects (rotation and magnetic field in the radiative zone, rotation and magnetic field in the convective zone) in the stellar evolution code. The hierarchy of all these effects will be useful and the formalism we have developed is totally general.

\section*{Acknowledgments}
The authors would like to thank A. S. Brun and T. Rashba for valuable discussions on the subject{, as well as the anonymous referee for her/his constructive remarks}.

\appendix
\section{The perturbation terms for the stellar structural modification}
We first expand the gravific potential around the non-magnetic state as
\begin{eqnarray}
\phi(r,\theta)&=&\phi_{0}\left(r\right)+\phi^{\left(1\right)}\left(r,\theta \right)\nonumber\\
&=&\phi_{0}(r)+\sum_{l\geqslant0}{\widehat\phi}_{l}(r)P_{l}\left(\cos\theta\right)
\end{eqnarray}
where $\phi_{0}(r)=- G M(r) / r $ is the potential of the non-magnetic star, $\widehat{\phi}_{l}$ the fluctuation for the mode $l$ due to the magnetic field and $P_{l}\left(\cos\theta\right)$ is the associated Legendre polynomial in the axisymmetric case.
Likewise, we expand the density $\rho$ and the pressure $P$ as:
\begin{eqnarray}
\rho\left(r,\theta \right) &=& \rho_{0}\left(r\right)+\rho^{\left(1\right)}\left(r,\theta \right)
 \nonumber\\
 &=& \rho_{0}\left(r\right)+\sum_{l \geqslant 0}\widehat{\rho}_{l}\left(r \right)P_{l}\left(\cos\theta\right);\\
P\left(r,\theta\right)
&=& P_{0}\left(r\right)+P^{\left(1\right)}\left(r,\theta\right)\nonumber \\
&=& P_{0}\left(r\right)+\sum_{l \geqslant 0}\widehat{P}_{l}\left(r\right)P_{l}\left(\cos\theta\right).
\end{eqnarray} 
We now consider the self-gravitating hydrostatic balance which is ruled by   
\begin{equation}
\frac{\bnab P}{\rho} = -\bnab \phi + \frac{\FLor}{\rho}
\end{equation}
and the Poisson's equation
\begin{equation}
\nabla^2\Phi=4\pi G\rho.
\end{equation} 
Expanded to the first order, these leads to: 
\begin{equation}
\bnab P^{\left(0\right)}=-\rho_{0}\bnab\phi_{0}
\end{equation}
\begin{equation}
\nabla^2\Phi^{\left(0\right)}=4\pi G\rho_{0}
\end{equation} 
and to
\begin{equation}
\bnab P^{\left(1\right)}=-\rho_{0}\bnab\phi^{\left(1\right)}-\rho^{\left(1\right)}\bnab\phi_{0}+\FLor
\label{Per1}
\end{equation}
\begin{equation}
\nabla^2\Phi^{\left(1\right)}=4\pi G\rho_{1}.
\end{equation}   
The pressure fluctuation is eliminated by taking the curl of Eq. (\ref{Per1}), which gives:  
\begin{eqnarray}
\PA{\rho^{\left(1\right)}}{\theta} = \frac{1}{g_{0}} \left[ 
\PA{\rho_{0}}{r} \PA{\phi^{\left(1\right)}}{\theta} +
\PA{F_{\mathcal{L},r}}{\theta} - 
\PA{\left(r F_{\mathcal{L}, \theta}\right)}{r} \right]
\label{pertrho}
\end{eqnarray}
where $g_{0} = - {\rm d}\Phi_{0}/{\rm d}r$. Next we insert the modal expansion  of $\rho^{\left(1\right)}$ and those of the components of the Lorentz force:
\begin{eqnarray}
{F}_{{\mathcal L},r}\left(r,\theta\right) &=& \sum_{l}\mathcal{X}_{\mbf{F}\!\!_{\mathcal{L}};l} \left(r\right)P_{l}\left(\cos\theta\right),\\
{\ F}_{{\mathcal L},\theta}\left(r,\theta\right) &=& -\sum_{l}\mathcal{Y}_{\mbf{F}\!\!_{\mathcal{L}};l}\left(r\right)\partial_{\theta}P_{l}\left(\cos\theta\right).\;\;\;\;\;
\end{eqnarray}  
After integration in $\theta$, this yields to the modal amplitude of the density perturbation around the non-magnetic state:
\begin{equation}
\widehat{\rho}_{l}=\frac{1}{g_{0}} \left[ \frac{{\rm d}\rho_{0}}{{\rm d}r} \widehat{\phi}_{l}+\mathcal{X}_{\mbf{F}\!\!_{\mathcal{L}};l}+\frac{{\rm d}}{{\rm d}r}\left(r\mathcal{Y}_{\mbf{F}\!\!_{\mathcal{L}};l}\right)\right] .
\label{rhol}
\end{equation}  
It remains to insert this expression in the perturbed Poisson equation $\nabla^2 \widehat{\phi}_{l} =
 4 \pi G \widehat{\rho}_{l}$ to retrieve Sweet's result
\begin{eqnarray}
\frac{1}{r}\frac{{\rm d}^2}{{\rm d}r^2}\left(r\widehat{\phi}_{l}\right)-\frac{l(l+1)}{r^2}\widehat{\phi}_{l}-\frac{4\pi G}{g_{0}}\frac{{\rm d}\rho_{0}}{{\rm d}r}\widehat{\phi}_{l}\nonumber \\
=\frac{4\pi G}{g_{0}} \left[\mathcal{X}_{\mbf{F}\!\!_{\mathcal{L}};l}+\frac{{\rm d}}{{\rm d}r}\left(r\mathcal{Y}_{\mbf{F}\!\!_{\mathcal{L}};l}\right)\right] .
\label{poissonpert}
\end{eqnarray}
The boundary conditions applied to (\ref{poissonpert}) are:
\begin{equation}
\widehat{\phi}_{0}=K\quad\hbox{and}\quad\widehat{\phi}_{l>0}=0\\ 
\end{equation}
at the center ($r=0$), where $K$ is a real, and
\begin{equation}
\frac{\rm d}{{\rm d}r}\widehat{\phi}_{l}-\frac{\left(l+1\right)}{r}\widehat{\phi}_{l}=0
\label{bc}
\end{equation}
at the surface that corresponds to the continuity of the gravific potential with the external multipolar one (for $r \ge R_{*}$) which can be expressed as in Roxburgh (2001):
\begin{equation}
\phi \!=\!-\frac{G M_{*}}{r}\left[1-\sum_{l\geqslant0}J_{l}\left(\frac{R_*}{r}\right)^{l+1}P_{l}\left(\cos\theta\right)\right]\!\!
\label{Multipole}
\end{equation}
where $M_{*}$ is the stellar mass and $J_{l}$ are the gravitational multipolar moment for the mode $l$, which are therefore the gravitational potential perturbations at the surface. 
The perturbation in pressure $\widehat{P}_{l}$ is then obtained from the  
$\theta$-component of the hydrostatic equation, which gives
\begin{equation}
\widehat{P}_{l}= - \rho_{0} \widehat{\phi}_{l} - r \mathcal{Y}_{\mbf{F}\!\!_{\mathcal L};l}.
\label{Pl}
\end{equation}
Finally, diagnosis from the stellar radius variation induced by the magnetic field can be established. Beginning with the definition of the radius of an isobar given by:
\begin{equation}
r_P(r, \theta) = r \left[1 + \sum_{l\geqslant0}c_{l}(r) P_{l}(\cos\theta)\right]
\end{equation}
we can identify using Eq. (\ref{Pl}) as in Mathis \& Zahn 2004 (cf. Eqs. 4 \& 6 in this paper):
\begin{equation}
c_{l}=-\frac{1}{r}\frac{\widehat{P}_{l}} {{{\rm d} P_{0}/{\rm d}r}}=\frac{\rho_{0}}{{{\rm d} P_{0}/{\rm d}r}}\left( \frac{1}{r} \widehat{\phi}_{l}+\frac{\mathcal{Y}_{\mbf{F}\!\!_{\mathcal L};l}}{\rho_{0}}\right).
\label{cl}
\end{equation}
\section{{Magnetic Equilibrium Configuration}}
{
Following the work by \cite{woltjer59}, the following Grad-Shafranov equation is obtained in the case of zero-torque Lorentz force per unit mass $\mbf{F}_{\mathcal L}/\rho$ : 
\begin{equation}
\Delta^{*}\Psi+F\left(\Psi\right)\partial_{\Psi}\left[F\left(\Psi\right)\right]=-\mu_0 r^2 \sin^2\theta\,\overline\rho\, G\left(\Psi\right).
\label{GS}
\end{equation}
Where $F$ and $G$ are arbitrary functions to be determined. Assuming that these quantities are regular, they can be expanded in the general way according to 
\begin{eqnarray}
F(\Psi) &=& \sum_{i=0}^{\infty}\frac{\lambda_{i}}{R} \Psi^{i};\\
G\left(\Psi\right)&=&\sum_{j=0}^{\infty}\beta_{j} \Psi^{j}.
\end{eqnarray}
Then, Eq. (\ref{GS}) becomes:
\begin{eqnarray}
\Delta^{*}\Psi+\sum_{k>0}\frac{\Lambda_{k}}{R^2}\Psi^{k}=-\mu_0 r^2 \sin^2\theta\,\rho\, \sum_{j=0}^{\infty}\beta_{j} \Psi^{j}, 
\label{GSNL}
\end{eqnarray}
where $\Lambda_k=\sum_{i_1>0}\sum_{i_2>0}\left\{i_2 \lambda_{i_1}\lambda_{i_2}\delta_{i_1+i_2-1,k}\right\}$, $\delta$ being the usual Kronecker symbol. 
This is the generalization of the Grad-Shafranov-type equation obtained by Prendergast (1956) for the compressible states.\\
Looking at the simplest linear solution non-singular for the field at the origin, this equation takes on the form\footnote{It can be shown that this linear equation arises from the conservation of the invariants of the problem which are the global helicity, the mass encompassed in potential flux surfaces and the total mass of the considered stellar region
.}
\begin{equation}
\Delta^{*}\Psi +\frac{\lambda_1^2}{R^2}\, \Psi = - \mu_0\,\rho\, r^2 \sin^2 \theta\,\beta_0,
\label{GSDM}
\end{equation}
whose solution can be found using Green's function method for the specified set of boundary conditions.}
\end{document}